\documentclass[aps,prd,twocolumn,groupedaddress,amssymb]{revtex4-1}
\usepackage{amsmath}
\usepackage{graphicx}
\usepackage{xcolor,footnote} 
\DeclareMathOperator{\Tr}{Tr}
\DeclareMathOperator{\B}{B}
\DeclareMathOperator{\R}{Re}
\DeclareMathOperator{\sign}{sign}

\begin{document}

\title{Dual representation of lattice QCD with \\
worldlines and worldsheets of abelian color fluxes}

\author{Carlotta Marchis}
\email[]{carla.marchis@uni-graz.at}
\author{Christof Gattringer}
\email[]{christof.gattringer@uni-graz.at}

\affiliation{Graz Universit\"{a}t, Institut f\"{u}r Physik, Universit\"{a}tsplatz 5, 8010 Graz, Austria}

\date{20.12.2017}

\begin{abstract}
We present a new dual representation for lattice QCD in terms of wordlines and worldsheets. The exact reformulation 
is carried out using the recently developed abelian color flux method where the action is decomposed into 
commuting minimal terms that connect different colors on neighboring sites. Expanding the Boltzmann factors for these
commuting terms allows one to reorganize the gauge field contributions according to links such that the gauge fields
can be integrated out in closed form. The emerging constraints give the dual variables the structure of worldlines for 
the fermions and worldsheets for the gauge degrees of freedom. The partition sum has the form of a strong coupling
expansion and with the abelian color flux approach discussed here all coefficients of the expansion are known in 
closed form. We present the dual form for three cases: pure SU(3) lattice gauge theory, strong coupling QCD and full 
QCD, and discuss in detail the constraints for the color fluxes and their physical interpretation.
\end{abstract}

\maketitle

\section{Introduction}
An important strategy in theoretical physics is to find different representations of a system, such that after rewriting
a model in terms of new degrees of freedom different physical aspects are revealed or new methods 
can be applied. In the context of lattice field theories exact transformations to representations in terms of worldlines
for matter fields and worldsheets for gauge degrees of freedom have been studied in recent years 
(see, e.g., the reviews at the annual lattice conferences 
\cite{Chandrasekharan:2008gp,deForcrand:2010ys,Wolff:2010zu,Gattringer:2014nxa}). 
A strong motivation for this line of work is the sign problem at finite chemical potential, which in some
models can be overcome with worldline/worldsheet representations, such that finite density simulations 
become accessible. However, recently also more abstract questions were addressed concerning 
the form of the constraints for the dual variables (i.e., worldlines and worldsheets) for different symmetries 
of the conventional representation.  

Finding dual representations for theories with abelian symmetries is essentially a closed case, see, e.g., the 
standard review \cite{Savit:1979ny}. For many abelian systems a second transformation to yet another set of variables 
allows one to solve all constraints and to arrive at a completely dual form in the Kramers-Wannier sense 
\cite{Kramers:1941zz}. However, for non-abelian symmetries the situation is far less advanced.  The key technique 
is strong coupling expansion which has been explored since the earliest days of lattice field theory 
\cite{Wilson:1974sk,Drouffe:1983fv,Rossi:1984cv,Dagotto:1986xt,Karsch:1988zx}. Recently, diagrammatic
representations in terms of worldlines and worldsheets for QCD and QCD-like lattice field theories have seen 
a prominent revival, see, e.g., \cite{Chandrasekharan:2005dn,deForcrand:2009dh,Unger:2011it,deForcrand:2014tha,
Vairinhos:2014uxa,Leme:2017fgn,deForcrand:2017fky,Bruckmann:2017vri,Borisenko:2017gql,Gattringer:2016lml,Marchis:2016cpe,Gattringer:2017hhn}, mostly driven by the quest for finding new 
representations to solve the aforementioned sign problem of QCD. However, so far no real and positive 
finite density representations were found (including the approach presented here) and obviously new concepts,
such as partial resummations are needed for possible applications in finite density simulations.

In this paper we present a dual representation of lattice QCD in terms of worldlines and worldsheets based on the
recently introduced {\sl ''abelian color flux (ACF) approach''} 
\cite{Gattringer:2016lml,Marchis:2016cpe,Gattringer:2017hhn}. While most approaches to strong coupling
representations of non-abelian theories rely on group integrals, often in the form of character expansion, the 
ACF approach decomposes the action into its smallest possible units, which are terms that connect different color
indices on neighboring sites of the lattice. These objects are either complex numbers for the gauge field
action or Grassmann bilinears for the fermions and thus commute in both cases. After expanding the 
individual Boltzmann factors one can reorder all terms and organize them 
with respect to links, such that they can be integrated over with the link-based Haar measure. No long range
interdependencies of the integrals emerge and all terms of the ACF form of the strong coupling expansion are obtained
as closed expressions. We stress at this point that some of the weights have negative sign, so without some form 
of resummation our representation cannot be directly used in a Monte Carlo simulation.

In this paper we focus on working out the ACF formulation for QCD, starting with the simpler cases of pure SU(3) gauge 
theory and strong coupling QCD, and deriving from those two limiting cases the full dual form of lattice QCD in terms
of worldlines and worldsheets. We discuss in detail the form of the constraints that emerge for our dual degrees of 
freedom, and show that they have the form of a conservation law for fluxes of all three colors ({\sl ''color conservation 
constraints''}) and a second set of constraints that ensure the equal distribution of flux among the colors ({\sl ''color
exchange constraints''}). We discuss the implications and geometrical interpretation of the constraints for all three cases 
we consider, i.e., pure SU(3) gauge theory, strong coupling QCD and full QCD. For the case of strong coupling 
QCD we discuss the behavior of the strong coupling baryon loops and show that they are closely related to free 
staggered fermions for the baryons, embedded in a background of local fermion monomials with positive weights.

\section{SU(3) lattice gauge theory \label{su3gauge}}

We start the presentation with deriving the worldsheet representation for pure SU(3) lattice gauge theory. 
We work with the Wilson gauge action
\begin{equation}
	S_G[U] \; = \; -\dfrac{\beta}{3} \sum_{x,\mu < \nu} 
	\R \Tr U_{x,\mu} \; U_{x+\hat{\mu},\nu} \, U_{x+\hat{\nu},\mu}^{\dagger} \, U_{x,\nu}^\dagger \; ,
	\label{eq:actionsu3}
\end{equation}
where $U_{x,\mu} \in$ SU(3) are the dynamical degrees of freedom of the theory. They live on the links $(x,\mu)$ 
of a four-dimensional lattice with periodic boundary conditions. The size of the lattice, i.e., the number of sites will be
denoted by $V$. The partition function $Z$ is obtained by integrating 
the Boltzmann factor $e^{-S_G[U]}$ with the product of SU(3) Haar measures 
$\int \! D[U] = \prod_{x,\mu} \int_{\text{SU(3)}} dU_{x,\mu}$, 
\begin{equation}
	Z = \int \! D[U] \; e^{-S_G[U]} \, .
	\label{eq:partitionsumsu3}
\end{equation}

As already outlined in the introduction, the first step of our approach consists of writing explicitly the 
trace and the matrix multiplications in the action (\ref{eq:actionsu3}) as sums over color indices for products 
of gauge link elements $U_{x,\mu}^{ab}$,
\begin{align}
	\nonumber
	S_G[U] \; = \; -\dfrac{\beta}{6} \sum_{x,\mu < \nu} & \; \sum_{a,b,c,d=1}^{3} \!\! \Big[
	U_{x,\mu}^{ab} U_{x+\hat{\mu},\nu}^{bc} U_{x+\hat{\nu},\mu}^{dc \ \star} U_{x,\nu}^{ad \ \star} \\
	& \; + \;U_{x,\mu}^{ab \ \star} U_{x+\hat{\mu},\nu}^{bc \ \star} U_{x+\hat{\nu},\mu}^{dc} U_{x,\nu}^{ad} \Big]\, .
	\label{eq:abelianactionsu3}
\end{align} 
The two products $U_{x,\mu}^{ab} U_{x+\hat{\mu},\nu}^{bc} U_{x+\hat{\nu},\mu}^{dc \ \star} U_{x,\nu}^{ad \ \star}$ and 
$U_{x,\mu}^{ab \ \star} U_{x+\hat{\mu},\nu}^{bc \ \star} U_{x+\hat{\nu},\mu}^{dc} U_{x,\nu}^{ad}$ 
are the objects we refer to as the {\sl ''Abelian Color Cycles''} (ACCs) \cite{Gattringer:2016lml}. 
They are products of complex numbers
and can be interpreted as paths in color space closing around plaquettes. 
In space-time we label the plaquettes $(x,\mu\nu)$
with the site $x$ in their lower left corner and the two directions $\mu < \nu$. 
The labelling of the ACCs is then completed 
by providing the values $(a,b,c,d)$ of the color indices at the four corners 
of the plaquette which determine the path in 
color space. 

\begin{figure}
	\vspace*{0.5cm}
	\includegraphics[width=6cm]{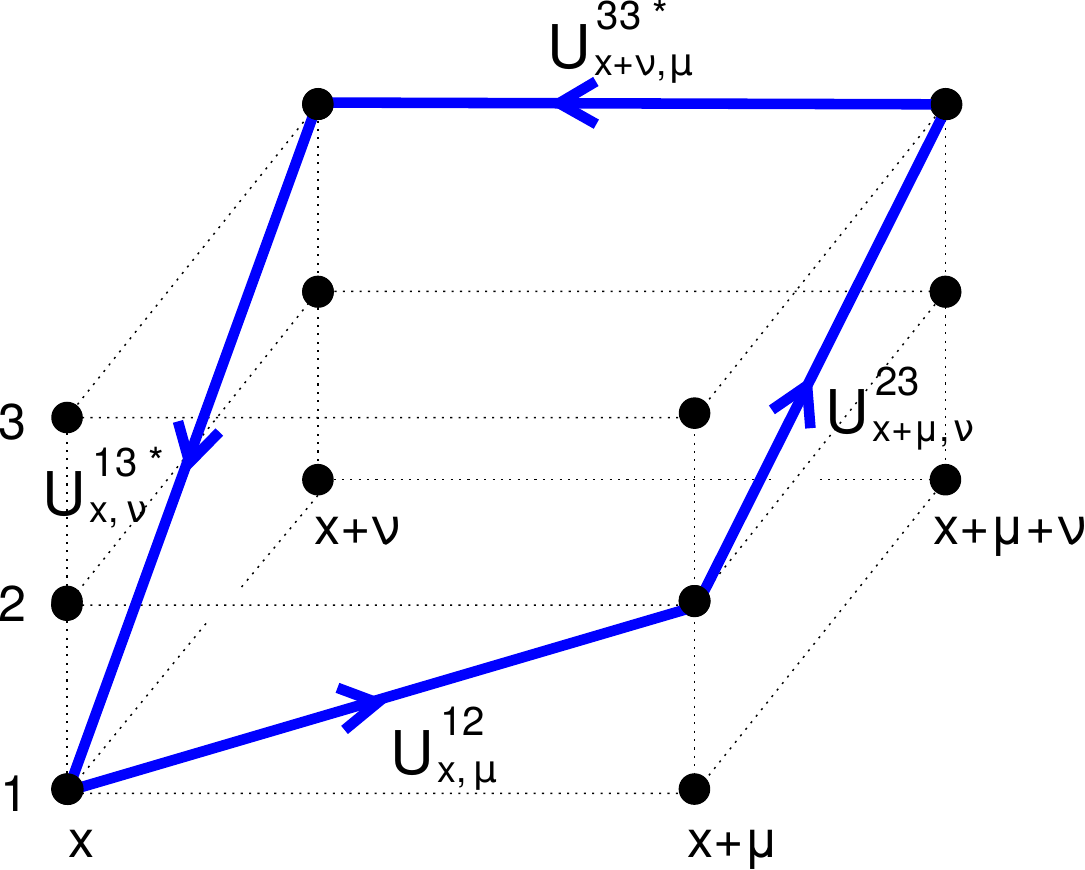}%
	\caption{Graphical representation of the $(1,2,3,3)$-ACC, which explicitly is given by 
	$U_{x,\mu}^{12} U_{x + \hat{\mu},\nu}^{23} U_{x + \hat{\nu},\mu}^{33 \ \star} U_{x,\nu}^{13 \ \star}$. 
	This ACC closes around the plaquette $(x,\mu\nu)$ running through the sequence 
	$(1,2,3,3)$ of color indices at the corners of the plaquette. In the graphical representation the color degrees of 
	 freedom are shown as three distinct layers of the space-time lattice, labelled with 1,2, and 3 on the lhs.\ of the plot. 
	 Each of the matrix elements $U_{x,\mu}^{ab}$ constituting the ACC is 
	 represented by an arrow along the corresponding link $(x,\mu)$ connecting color $a$ with color $b$. For complex
	 conjugate matrix elements the link is run through with negative orientation.}
	\label{fig:plaquette_su3}
\end{figure}

To give an example, in Fig.~\ref{fig:plaquette_su3} we graphically illustrate 
the $(1,2,3,3)$-ACC which in explicit form is given by
$U_{x,\mu}^{12} U_{x+\hat{\mu},\nu}^{23} U_{x+\hat{\nu},\mu}^{33 \ \star} U_{x,\nu}^{13 \ \star}$. 
The color degrees of freedom are represented by using a lattice with three layers, which 
in the figure we sketch in light grey as three copies of the plaquette we consider.  The terms in the $(1,2,3,3)$-ACC 
then all have a simple graphical representation: The factor $U_{x,\mu}^{12}$ is represented by 
an arrow connecting the color index 1 at $x$ with the color index $2$ 
at $x + \hat{\mu}$. The factor $U_{x+\hat{\mu},\nu}^{23}$
then continues from color 2 to color 3 along the link from $x+\hat{\mu}$ to $x+\hat{\mu} + \hat{\nu}$. Link matrix elements
with a complex conjugation are interpreted as running in negative direction, such that $U_{x+\hat{\nu},\mu}^{33 \ \star}$
leads from color 3 at $x+\hat{\mu} + \hat{\nu}$ to color 3 at $x+\hat{\nu}$ and $U_{x,\nu}^{13 \ \star}$ closes the loop 
leading from color 3 at $x+\hat{\nu}$ to color 1 at $x$. 

The rule of reverting the orientation with complex conjugation implies that the ACCs in the second summand of 
(\ref{eq:abelianactionsu3}) run around the plaquette $(x,\mu\nu)$ with mathematically negative orientation. 
Since for SU(3) there are three different possible choices for the color at every corner of the plaquette there is a total 
of  $3^4=81$ different ACCs, each contributing with both orientations to (\ref{eq:abelianactionsu3}).

The ACC decomposition (\ref{eq:abelianactionsu3}) of the action (\ref{eq:actionsu3}) allows us to completely 
factorize the Boltzmann factor and to proceed with the dualization of the theory along the lines of the construction for abelian gauge fields 
(see, e.g., \cite{Mercado:2013ola,Mercado:2013yta}): 

\begin{align}
\label{eq:partitionsumsu32}
&Z  = \int \! \!\! D[U] \!\! \prod_{x,\mu<\nu} \, \prod_{a,b,c,d} \!\!
e^{\frac{\beta}{6}\left( U_{x,\mu}^{ab} U_{x+\hat{\mu},\nu}^{bc} U_{x+\hat{\nu},\mu}^{dc \ \star} U_{x,\nu}^{ad \ \star} + \; c.c.\right)} 
\nonumber  \\
& = \int \!\!\! D[U] \!\!
\prod_{x,\mu<\nu} \; \prod_{a,b,c,d} \, \sum_{n_{x,\mu\nu}^{abcd} = 0}^{\infty} \; \sum_{\overline{n}_{x,\mu\nu}^{abcd} = 0}^{\infty} \!
\dfrac{ \left( \beta/6 \right)^{n_{x,\mu\nu}^{abcd} + \overline{n}_{x,\mu\nu}^{abcd}}}{n_{x,\mu\nu}^{abcd}\, ! \; \; 
\overline{n}_{x,\mu\nu}^{abcd}\, !} \nonumber \\
& \qquad  \times 
\left( U_{x,\mu}^{ab} U_{x+\hat{\mu},\nu}^{bc} U_{x+\hat{\nu},\mu}^{dc \ \star} 
U_{x,\nu}^{ad \ \star} \right)^{n_{x,\mu\nu}^{abcd}}
\left( c. c. \right)^{\overline{n}_{x,\mu\nu}^{abcd}} .
\end{align}
In the first step we have written all sums in the exponents as products over the individual  Boltzmann weights  
$e^{\frac{\beta}{6} U_{x,\mu}^{ab} U_{x+\hat{\mu},\nu}^{bc} U_{x+\hat{\nu},\mu}^{dc \ \star} U_{x,\nu}^{ad \ \star}}$ and 
$e^{\frac{\beta}{6} U_{x,\mu}^{ab \ \star} U_{x+\hat{\mu},\nu}^{bc \ \star} U_{x+\hat{\nu},\mu}^{dc} U_{x,\nu}^{ad}}$ 
for the ACCs with positive and negative orientation. In the second step we expand each factor in a 
Taylor series, thus introducing two sets of expansion indices assigned to the plaquettes: 
$n_{x,\mu\nu}^{abcd} \in \mathbb{N}_{0}$ and $\overline{n}_{x,\mu\nu}^{abcd} \in \mathbb{N}_{0}$ 
where the color indices $a,b,c$ and $d$ each can have the values 1,2 or 3. The variables $n_{x,\mu\nu}^{abcd}$
correspond to the units of flux with color indices $a,b,c,d$ around the plaquette $(x,\mu\nu)$
in positive orientation and $\overline{n}_{x,\mu\nu}^{abcd}$ is used for flux with negative orientation. 

All the factors in the sums in (\ref{eq:partitionsumsu32}) are products of 
complex numbers, such that we can freely commute them and reorganize 
them as in the abelian case. Ordering the terms with respect to the links $(x,\mu)$ where we will subsequently integrate them 
with the Haar measure, the partition sum assumes the form
\begin{align}
\label{eq:partitionsumsu33}
Z \; = \; \sum_{\{n, \overline{n}\}} & \bigg[ \prod_{x,\mu<\nu} \prod_{a,b,c,d}  
\dfrac{ \left( \beta/6 \right)^{n_{x,\mu\nu}^{abcd} + \overline{n}_{x,\mu\nu}^{abcd}}}{n_{x,\mu\nu}^{abcd}\, !  \; \; 
\overline{n}_{x,\mu\nu}^{abcd}\, !} \bigg] \\
&\times \bigg[\prod_{x,\mu} \int \! \! dU_{x,\mu} \; \prod_{a,b} \left( U_{x,\mu}^{ab} \right) ^{N_{x,\mu}^{ab}} 
\left( U_{x,\mu}^{ab \ \star} \right)^{\overline{N}_{x,\mu}^{ab}} \bigg]\; ,
\nonumber
\end{align}
where for the sum over configurations of the variables 
$n_{x,\mu\nu}^{abcd}, \overline{n}_{x,\mu\nu}^{abcd} \in \mathbb{N}_{0}$  
we introduced the short hand notation 
$\sum_{\{n, \overline{n}\}} = \prod_{x,\mu<\nu} \prod_{a,b,c,d = 1}^{3} \sum_{n_{x,\mu\nu}^{abcd} = 0}^{\infty} 
\sum_{\overline{n}_{x,\mu\nu}^{abcd} = 0}^{\infty}$. The integer valued powers $N_{x,\mu\nu}^{abcd}$ and 
$\overline{N}_{x,\mu\nu}^{abcd}$ collect all $n_{x,\mu\nu}^{abcd}$ and $\overline{n}_{x,\mu\nu}^{abcd}$ where the 
matrix elements $U_{x,\mu}^{ab}$ and $U_{x,\mu}^{ab \ \star}$ appear. Explicitly they are given by
\begin{gather} 
\label{eq:powerN}
N_{x,\mu}^{ab}  = \!\! 
\sum_{\nu:\mu<\nu} \! n_{x,\mu\nu}^{abss} + \overline{n}_{x-\hat{\nu},\mu\nu}^{ssba} +\!\!
\sum_{\rho:\mu>\rho} \! \overline{n}_{x,\rho\mu}^{assb} + n_{x-\hat{\rho},\rho\mu}^{sabs},
\\
\label{eq:powerNbar}
\overline{N}_{x,\mu}^{ab} = \!\! 
\sum_{\nu:\mu<\nu} \! \overline{n}_{x,\mu\nu}^{abss} + n_{x-\hat{\nu},\mu\nu}^{ssba} + \!\!
\sum_{\rho:\mu>\rho} \! n_{x,\rho\mu}^{assb} + \overline{n}_{x-\hat{\rho},\rho\mu}^{sabs} .
\end{gather} 
The label $s$ introduced here is the short hand notation for an 
independent summation over all color indices replaced by $s$, 
e.g., $n_{x,\rho\mu}^{assb} \equiv \sum_{c,d} n_{x,\rho\mu}^{acdb}$.

The Haar measure integration in (\ref{eq:partitionsumsu33}) is now done using an explicit parametrization 
for the SU(3) matrices \cite{Bronzan:1988wa}:
\begin{widetext}
	\begin{equation}
	\label{eq:parametrization}
	U_{x,\mu} =\left(
	\begin{array}{ccc}
	c_1 c_2 \, e^{i\phi_1}  & s_1 \, e^{i\phi_3} & c_1 s_2 \, e^{i\phi_4}\\
	s_2 s_3 \, e^{-i\phi_4 -i\phi_5} - s_1 c_2 c_3 \, e^{i\phi_1 +i\phi_2 -i\phi_3} & 
	c_1 c_3 \, e^{i\phi_2} & - c_2 s_3 \, e^{-i\phi_1 -i\phi_5} - s_1 s_2 c_3 \, e^{i\phi_2 -i\phi_3 +i\phi_4}\\
	- s_2 c_3 \, e^{-i\phi_2 -i\phi_4} - s_1 c_2 s_3 \, e^{i\phi_1 -i\phi_3 +i\phi_5} & 
	c_1 s_3 \, e^{i\phi_5} & c_2 c_3 \, e^{-i\phi_1 -i\phi_2} - s_1 s_2 s_3 \, e^{-i\phi_3 +i\phi_4 +i\phi_5}
	\end{array} \right) \; .
	\end{equation}
\end{widetext}
The parameterization uses three angles $\theta_{x,\mu}^{(j)} \in [0, \pi/2]$, $j = 1,2,3$ 
and five phases  $\phi_{x,\mu}^{(j)} \in [-\pi,\pi]$, 
$j = 1\, ... \, 5$. 
In (\ref{eq:parametrization}) we use the abbreviations $c_{j} = \cos \theta_{x,\mu}^{(j)}$, 
$s_{j} = \sin \theta_{x,\mu}^{(j)}$ and $\phi_{j} = \phi_{x,\mu}^{(j)}$. 
For the parameterization (\ref{eq:parametrization}) the normalized Haar measure is given by
\begin{equation}
dU_{x,\mu}  \; = \; 16 \,  
d\theta_1 c_1^3 s_1 \; d\theta_2 c_2 s_2 \; d\theta_3 c_3 s_3 \, \prod_{j=1}^5 \frac{d\phi_j}{ 2\pi} \, .
\label{eq:haarmeasure}
\end{equation}
We will see below that the integration over the angles $\phi_{x,\mu}^{(j)}, \, j = 1, 2 \, ... \, 5$ 
will give rise to constraints for the variables $n_{x,\mu\nu}^{abcd}$ and $\overline{n}_{x,\mu\nu}^{abcd}$. 
In order to give these constraints a transparent form it is useful to perform the change of variables
\begin{gather}
\label{eq:cycleoccupationnumbers}
n_{x,\mu\nu}^{abcd} - \overline{n}_{x,\mu\nu}^{abcd} = p_{x,\mu\nu}^{abcd} \quad , \quad 
p_{x,\mu\nu}^{abcd} \in \mathbb{Z} \; ,\\
\label{eq:auxiliary}
n_{x,\mu\nu}^{abcd} + \overline{n}_{x,\mu\nu}^{abcd} = 
|p_{x,\mu\nu}^{abcd}| + 2 l_{x,\mu\nu}^{abcd} \quad , \quad l_{x,\mu\nu}^{abcd} \in \mathbb{N}_0 \; ,
\end{gather}
and instead of summing over the configurations of the $n_{x,\mu\nu}^{abcd}$ and 
$\overline{n}_{x,\mu\nu}^{abcd}$ to sum over configurations of the $p_{x,\mu\nu}^{abcd}$ and $l_{x,\mu\nu}^{abcd}$. 
The sets of variables $p_{x,\mu\nu}^{abcd} \in \mathbb{Z}$ and $l_{x,\mu\nu}^{abcd} \in \mathbb{N}_0$, which are both 
assigned to the plaquettes of the lattice will be the new dynamical dual degrees of freedom that we use in the 
partition sum after integrating out the conventional fields $U_{x,\mu}$. 

The $p_{x,\mu\nu}^{abcd}$ will be subject to constraints and for understanding 
these constraints it is important to discuss the geometrical interpretation of the $p_{x,\mu\nu}^{abcd}$:
From the definition (\ref{eq:cycleoccupationnumbers}) and the interpretation 
of the $n_{x,\mu\nu}^{abcd}$ ($\overline{n}_{x,\mu\nu}^{abcd}$) as the activation 
numbers for $(a,b,c,d)$-ACCs with positive (negative) 
orientation it is clear that the new variables $p_{x,\mu\nu}^{abcd}$ activate 
$|p_{x,\mu\nu}^{abcd}|$ units of flux for the $(a,b,c,d)$-ACC 
on the plaquette $(x,\mu\nu)$, with the orientation of the flux given by the sign of the 
$p_{x,\mu\nu}^{abcd}$. We refer to the
$p_{x,\mu\nu}^{abcd}$ as  \textit{''cycle occupation numbers''}. 
The $l_{x,\mu\nu}^{abcd}$ are not subject to constraints and 
we simply refer to them as \textit{''auxiliary plaquette variables''}. 

For further simplification it is convenient to introduce the link fluxes
\begin{equation}
	\label{eq:Jfluxes}
	J_{x,\mu}^{ab} = \! \!
	\sum_{\nu:\mu<\nu} \! [\, p_{x,\mu\nu}^{abss} - p_{x-\hat{\nu},\mu\nu}^{ssba} \, ] - \!\!
	\sum_{\rho:\mu>\rho} \! [\, p_{x,\rho\mu}^{assb} - p_{x-\hat{\rho},\rho\mu}^{sabs} \, ]  ,
\end{equation}
and the auxiliary link sums	
\begin{align}
		S_{x,\mu}^{ab} &=  
		\sum_{\nu:\mu<\nu}  [ |p_{x,\mu\nu}^{abss}| + |p_{x-\hat{\nu},\mu\nu}^{ssba}| + 
		2(l_{x,\mu\nu}^{abss} + l_{x-\hat{\nu},\mu\nu}^{ssba}) ] 
		\nonumber \\
		&\, + \sum_{\rho:\mu>\rho} [ |p_{x,\rho\mu}^{assb}| + |p_{x-\hat{\rho},\rho\mu}^{sabs}| + 
		2(l_{x,\rho\mu}^{assb} + l_{x-\hat{\rho},\rho\mu}^{sabs}) ] \, .
		\label{eq:Sfluxes}
\end{align}
We will see that only the fluxes $J_{x,\mu}^{ab}$ will appear in the constraints and we thus have to 
extend our geometrical interpretation of the dual variables to these objects: $J_{x,\mu}^{ab}$ is the total 
flux from color $a$ on site $x$ to color $b$ on site $x + \hat{\mu}$. This flux receives contributions from 
all the ACCs that are attached to the link $(x,\mu)$ and that contain the path from color $a$ to $b$ along 
that link. So, if we consider the plaquette $(x,\mu\nu)$, with $\mu < \nu$, we have 9 different ACCs that 
contribute to that flux, namely the ones corresponding to the cycle occupation numbers $p_{x,\mu\nu}^{abef}$, 
where $a$ and $b$ are the color indices which we fix at $x$ and $x + \hat{\mu}$.
The colors $e$ and $f$ determine the ACC at the remaining two corners of $(x,\mu\nu)$. 
Both, $e$ and $f$ can be chosen independently from 
the set $\{1,2,3\}$ such that we have $3^2 = 9$ possibilities. Since the flux of the ACCs on the plaquette $(x,\mu\nu)$ 
has a positive orientation along the link $(x,\mu)$, these 9 ACCs contribute with a positive sign in the definition 
(\ref{eq:Jfluxes}) of the fluxes $J_{x,\mu}^{ab}$. However, $J_{x,\mu}^{ab}$ receives contributions from all 
plaquettes that contain the link $(x,\mu)$, such as the plaquettes $(x,\rho\mu)$ with $\rho < \mu$. 
For this case $J_{x,\mu}^{ab}$ receives contributions from the 9 ACCs with occupation numbers $p_{x,\rho\mu}^{aefb}$, 
but here the link $(x,\mu)$ is run through with negative orientation, such that the 
$p_{x,\rho\mu}^{aefb}$ contribute with a negative sign. For the remaining plaquettes that contain the link $(x,\mu)$
and thus contribute to $J_{x,\mu}^{ab}$ an analogous discussion holds.

\begin{figure}[b!!]
	\includegraphics[width=6.5cm]{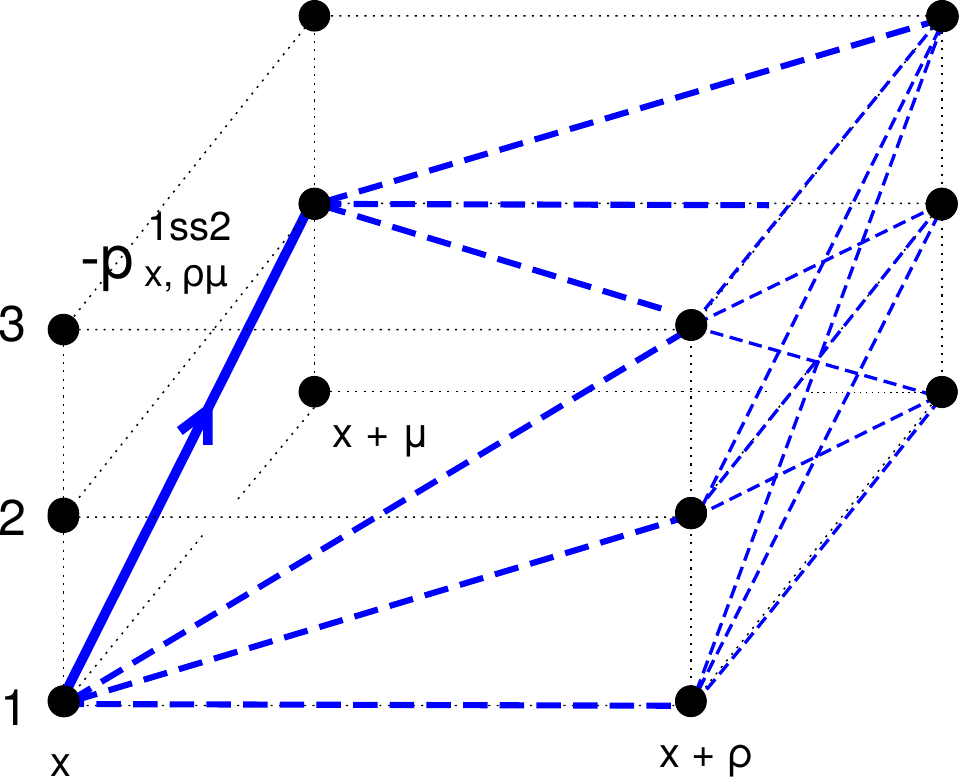}%
	\caption{Graphical illustration of the sum $- p_{x,\rho\mu}^{1ss2}$ contributing to the flux $J_{x,\mu}^{12}$.
	\label{fig:color_flux}}
\end{figure}

In order to illustrate the geometrical interpretation of the cycle occupation numbers that contribute to 
a given $J_{x,\mu}^{ab}$, in Fig.\ \ref{fig:color_flux} we illustrate the contributions from a 
plaquette $(x,\rho\mu)$ with $\rho < \mu$ to
$J_{x,\mu}^{12}$. The (1,2) flux on the link $(x,\mu)$ is fixed and represented with a full arrow pointing in the positive 
$\mu$ direction. The nine ACCs on the plaquette $(x,\rho\mu)$ 
that contribute to this flux are represented with dashed lines in the figure. 
Since the ACCs on the plaquette $(x,\rho\mu)$ have a negative orientation of the flux on the link $(x,\mu)$, 
they contribute with a negative sign to the flux $J_{x,\mu}^{12}$.

Having introduced the fluxes $J_{x,\mu}^{ab}$ and the auxiliary sums $S_{x,\mu}^{ab}$ we can rewrite the 
integers $N_{x,\mu}^{ab}$ and $\overline{N}_{x,\mu}^{ab}$ that denote the powers of 
$U_{x,\mu}^{ab}$ and $U_{x,\mu}^{ab \ \star}$ in (\ref{eq:partitionsumsu33})
in terms of the  $J_{x,\mu}^{ab}$ and $S_{x,\mu}^{ab}$ as
\begin{equation}
N_{x,\mu}^{ab} = \frac{S_{x,\mu}^{ab} + J_{x,\mu}^{ab}}{2} \; , \; \; 
	\label{eq:powerNbar2}
\overline{N}_{x,\mu}^{ab} = \frac{S_{x,\mu}^{ab} - J_{x,\mu}^{ab}}{2} \; .
\end{equation}
Since in (\ref{eq:partitionsumsu33}) the matrix elements $U_{x,\mu}^{ab}$ appear in the combination 
$\left( U_{x,\mu}^{ab} \right) ^{N_{x,\mu}^{ab}} \left( U_{x,\mu}^{ab \ \star} \right) ^{\overline{N}_{x,\mu}^{ab}}$
the form (\ref{eq:powerNbar2}) separates the moduli and  the phases of the matrix elements in a natural way. 

However, an additional step is still required before we arrive at the final form of the terms in (\ref{eq:partitionsumsu33}) 
where we can perform the Haar measure integration at each link. 
The problem is that some of the elements $U_{x,\mu}^{ab}$ of the matrix (\ref{eq:parametrization}) are not in the simple form 
$U_{x,\mu}^{ab} = r_{x,\mu}^{ab} e^{i \varphi_{x,\mu}^{ab}}$, but are sums 
$U_{x,\mu}^{ab} = \rho_{x,\mu}^{ab} e^{i \alpha_{x,\mu}^{ab}} + \omega_{x,\mu}^{ab} e^{i \beta_{x,\mu}^{ab}}$. 
More specifically, in the parameterization (\ref{eq:parametrization}) the $(a,b) = (2,1), (2,3), (3,1)$ and $(3,3)$ 
matrix elements are sums of two terms. For these entries we use the binomial theorem 
$(x + y)^{n} = \sum_{k = 0}^{n} \binom{n}{k} x^{n-k} y^{k}$ and rewrite their contribution 
in the integrand of (\ref{eq:partitionsumsu33}) as
\begin{eqnarray}
&&	\left( U_{x,\mu}^{ab} \right)^{N_{x,\mu}^{ab}} \!
	\left( U_{x,\mu}^{ab \ \star} \right)^{\overline{N}_{x,\mu}^{ab}} = 
	\nonumber \\ 	
&&
	\left(  \! \rho_{x,\mu}^{ab} e^{i \alpha_{x,\mu}^{ab}} \! + 
	\omega_{x,\mu}^{ab} e^{i \beta_{x,\mu}^{ab}} \! \right)^{\! N_{x,\mu}^{ab}} \!\!
	\left( \! \rho_{x,\mu}^{ab} e^{-i \alpha_{x,\mu}^{ab}} \! + 
	\omega_{x,\mu}^{ab} e^{-i \beta_{x,\mu}^{ab}} \! \right)^{\! \overline{N}_{x,\mu}^{ab}} 
\nonumber \\ 	
&& \; \hspace*{4mm} =	
	\!\! \sum_{m_{x,\mu}^{ab} = 0}^{N_{x,\mu}^{ab}} 
	\; \sum_{\overline{m}_{x,\mu}^{ab} = 0}^{\overline{N}_{x,\mu}^{ab}} \binom{N^{ab}_{x,\mu}}{m^{ab}_{x,\mu}}
	\binom{\overline{N}^{ab}_{x,\mu}}{\overline{m}^{ab}_{x,\mu}}
\nonumber \\ 	
&& \; \hspace*{4mm}
	\times \left( \rho_{x,\mu}^{ab} \right)^{s_{x,\mu}^{ab}} 
	\left( \omega_{x,\mu}^{ab} \right)^{S_{x,\mu}^{ab} - s_{x,\mu}^{ab}} 
	e^{i\alpha_{x,\mu}^{ab} j_{x,\mu}^{ab}} \, e^{i\beta_{x,\mu}^{ab} \left(J_{x,\mu}^{ab} - j_{x,\mu}^{ab}\right)}
\nonumber \\ 
&& \nonumber \\ 		
&& \; \hspace*{-1mm}
\mbox{with} \quad 
m_{x,\mu}^{ab} = 0,1 \, ... \; N_{x,\mu}^{ab} \; , \quad
\overline{m}_{x,\mu}^{ab} = 0,1 \, ... \; \overline{N}_{x,\mu}^{ab} \; ,
\nonumber \\ 
&& \; \hspace*{-1mm}
\mbox{and} 
\quad  \; j_{x,\mu}^{ab} \equiv m_{x,\mu}^{ab} - \overline{m}_{x,\mu}^{ab}\; , \; \; \; 
s_{x,\mu}^{ab} \equiv m_{x,\mu}^{ab} + \overline{m}_{x,\mu}^{ab} \; . \!\!\!\!
	\label{eq:binomial}
\end{eqnarray}
Note that the new auxiliary variables $m_{x,\mu}^{ab}$ and 
$\overline{m}_{x,\mu}^{ab}$ which we use for the binomial decomposition of the 
matrix elements with $(a,b) = (2,1), (2,3), (3,1)$ and $(3,3)$ live on the links of the lattice.

To obtain the final result for the partition function we substitute the parametrization 
(\ref{eq:parametrization}) and the Haar measure (\ref{eq:haarmeasure}) in (\ref{eq:partitionsumsu33}) 
and use the binomial decomposition (\ref{eq:binomial}). For the partition function we then obtain
\begin{widetext}
	\begin{align}
	\label{eq:partitionsumsu34}
	&Z \; = \; 2^{4V} \! \sum_{\{p,l\}} \sum_{\{m,\overline{m}\}} \! \Bigg[ \prod_{x,\mu<\nu} \prod_{a,b,c,d} \!
	\dfrac{ \left( \beta/2 \right)^{|p_{x,\mu\nu}^{abcd}| + 2\, l_{x,\mu\nu}^{abcd}}}{
	\left(|p_{x,\mu\nu}^{abcd}| + l_{x,\mu\nu}^{abcd}\right)! \; l_{x,\mu\nu}^{abcd}!} \Bigg] \!
	 \Bigg[ \prod_{x,\mu} (-1)^{S_{x,\mu}^{23} + S_{x,\mu}^{31} + s_{x,\mu}^{21} + s_{x,\mu}^{33}} \Bigg] \!
	\Bigg[ \prod_{x,\mu}  \prod_{a = 2,3}  \, \prod_{b=1,3} \!\!
	\binom{N_{x,\mu}^{ab}}{m_{x,\mu}^{ab}} \binom{\overline{N}_{x,\mu}^{ab}}{\overline{m}_{x,\mu}^{ab}} \!
	\Bigg]
	\nonumber
	\\
	& \hspace{6mm} \times \! \prod_{x,\mu}
	2 \! \int_{0}^{\pi/2} \!\!\!\!  d\theta_{x,\mu}^{(1)} \, 
	(\cos\theta_{x,\mu}^{(1)})^{3 + S_{x,\mu}^{11} + S_{x,\mu}^{13} + S_{x,\mu}^{22} + S_{x,\mu}^{32}}
	\; (\sin\theta_{x,\mu}^{(1)})^{1 + S_{x,\mu}^{12} + s_{x,\mu}^{21} +s_{x,\mu}^{23} + s_{x,\mu}^{31} + s_{x,\mu}^{33}} 
	\nonumber
	\\
	& \hspace{11mm} \times \!  
	2 \! \int_{0}^{\pi/2} \!\!\!\!  d\theta_{x,\mu}^{(2)} \, (\cos\theta_{x,\mu}^{(2)})^{1 + S_{x,\mu}^{11} + s_{x,\mu}^{21} +
	S_{x,\mu}^{23} - s_{x,\mu}^{23} + s_{x,\mu}^{31} + S_{x,\mu}^{33} - s_{x,\mu}^{33}} 
	\; (\sin\theta_{x,\mu}^{(2)})^{1 + S_{x,\mu}^{13} + S_{x,\mu}^{21} - s_{x,\mu}^{21} + s_{x,\mu}^{23} + 
	S_{x,\mu}^{31} - s_{x,\mu}^{31} + s_{x,\mu}^{33}} 
	\nonumber
	\\
	& \hspace{11mm} \times \! 
	2 \! \int_{0}^{\pi/2} \!\!\!\!  d\theta_{x,\mu}^{(3)} \, (\cos\theta_{x,\mu}^{(3)})^{1 + s_{x,\mu}^{21} + 
	S_{x,\mu}^{22} + s_{x,\mu}^{23} + S_{x,\mu}^{31} - s_{x,\mu}^{31} + S_{x,\mu}^{33} - s_{x,\mu}^{33}} 
	\; (\sin\theta_{x,\mu}^{(3)})^{1 + S_{x,\mu}^{21} - s_{x,\mu}^{21} + S_{x,\mu}^{23} - s_{x,\mu}^{23} + 
	s_{x,\mu}^{31} + S_{x,\mu}^{32} + s_{x,\mu}^{33}} 
	\nonumber
	\\
	&  \hspace{15mm} \times \! 
	\int_{0}^{2\pi} \! \dfrac{d\phi_{x,\mu}^{(1)}}{2\pi} \; e^{i\phi_{x,\mu}^{(1)}[J_{x,\mu}^{11}-
	J_{x,\mu}^{23}-J_{x,\mu}^{33}+j_{x,\mu}^{21}+j_{x,\mu}^{23}+j_{x,\mu}^{31}+j_{x,\mu}^{33}]}
	\int_{0}^{2\pi} \! \dfrac{d\phi_{x,\mu}^{(2)}}{2\pi} \; e^{i\phi_{x,\mu}^{(2)}[J_{x,\mu}^{22}-J_{x,\mu}^{31}-
	J_{x,\mu}^{33}+j_{x,\mu}^{21}+j_{x,\mu}^{23}+j_{x,\mu}^{31}+j_{x,\mu}^{33}]} \; 
	\nonumber
	\\
	&  \hspace{15mm} \times \! 
	\int_{0}^{2\pi} \! \dfrac{d\phi_{x,\mu}^{(3)}}{2\pi} \; e^{i\phi_{x,\mu}^{(3)}[J_{x,\mu}^{12}-j_{x,\mu}^{21}-
	j_{x,\mu}^{23}-j_{x,\mu}^{31}-j_{x,\mu}^{33}]}
	\int_{0}^{2\pi} \! \dfrac{d\phi_{x,\mu}^{(4)}}{2\pi} \; e^{i\phi_{x,\mu}^{(4)}[J_{x,\mu}^{13}-J_{x,\mu}^{21}-
	J_{x,\mu}^{31}+j_{x,\mu}^{21}+j_{x,\mu}^{23}+j_{x,\mu}^{31}+j_{x,\mu}^{33}]} \; 
	\nonumber
	\\
	&  \hspace{15mm} \times \! 
	\int_{0}^{2\pi} \! \dfrac{d\phi_{x,\mu}^{(5)}}{2\pi} \; e^{i\phi_{x,\mu}^{(5)}[J_{x,\mu}^{32}-J_{x,\mu}^{21}-
	J_{x,\mu}^{23}+j_{x,\mu}^{21}+j_{x,\mu}^{23}+j_{x,\mu}^{31}+j_{x,\mu}^{33}]} \; ,
	\end{align} 
\end{widetext}
where we introduced the short hand notation 
\begin{equation}
\sum_{\{p\}} = \! \! \prod_{x,\mu<\nu} \;  \prod_{a,b,c,d} \; \sum_{p_{x,\mu\nu}^{abcd} = -\infty}^{\infty} \!\!\!
, \; \; 
\sum_{\{l\}} = \! \! \prod_{x,\mu<\nu} \;  \prod_{a,b,c,d}  \; \sum_{l_{x,\mu\nu}^{abcd} = 0}^{\infty} ,
\end{equation}
for the sums over configurations of the cycle occupation numbers $p_{x,\mu\nu}^{abcd} \in \mathbb{Z}$ 
and the auxiliary plaquette variables $l_{x,\mu\nu}^{abcd} \in \mathbb{N}_{0}$,  as well as 
\begin{equation}
\sum_{\{m, \overline{m}\}} =  \; \prod_{x,\mu} \; \prod_{a = 2,3}  \; \prod_{b=1,3} \; 
\sum_{m_{x,\mu}^{ab} = 0}^{N_{x,\mu}^{ab}} \; \sum_{\overline{m}_{x,\mu}^{ab} = 0}^{\overline{N}_{x,\mu}^{ab}}  \; ,
\end{equation}
for the sums over configurations of the link based auxiliary variables $m_{x,\mu}^{ab}$ and $\overline{m}_{x,\mu}^{ab}$ 
used in the binomial decomposition (\ref{eq:binomial}).

A key step of our approach is that now, after expanding the Boltzmann factors for the individual ACCs and reorganizing 
all contributions with respect to links, in (\ref{eq:partitionsumsu34}) we can solve all Haar measure integrals in closed form. 
The integrals over the angles $\theta_{x,\mu}^{(j)}$ 
give rise to beta functions,
\begin{align}
\label{eq:betafunction}
2 \! \int_{0}^{\pi/2} \! \! \! \!\! d\theta (\cos \theta)^{n + 1} (\sin \theta)^{m + 1} = 
\B\left(\dfrac{n}{2} + 1\right| \left. \! \dfrac{m}{2} + 1\right) .
\end{align}
The integrals over the phase factors $\phi_{x,\mu}^{(j)}$ in (\ref{eq:partitionsumsu34}) give rise to Kronecker deltas 
(we use the notation $\delta(n) \equiv \delta_{n,0}$) which impose constraints on the dual variables.

Putting together all terms we can write the dual form of the partition function of pure SU(3) lattice gauge theory in the form
\begin{equation}
\label{eq:partitionsumsu35}
	Z = \sum_{\{p\}}  W_G[p] \; C_G[p]  \; ,
\end{equation}
where we have defined the link-based gauge constraints $C_G[p]$ that are given by 
\begin{eqnarray}
\label{linkconstraints}
C_{G} [p]  & = &  \prod_{x,\nu}  \delta(J_{x,\mu}^{12} + J_{x,\mu}^{13} - J_{x,\mu}^{21} - J_{x,\mu}^{31}) 
\\
&& \hspace{1.2mm} \times  \, 
\delta(J_{x,\mu}^{21} + J_{x,\mu}^{23} - J_{x,\mu}^{12} - J_{x,\mu}^{32}) 
\nonumber \\ 
&& \hspace{1.2mm} \times \, 
\delta(J_{x,\mu}^{11} + J_{x,\mu}^{12} - J_{x,\mu}^{23} - J_{x,\mu}^{33})
\nonumber \\ 
&& \hspace{1.2mm} \times \, 
\delta(J_{x,\mu}^{31} + J_{x,\mu}^{33} - J_{x,\mu}^{12} - J_{x,\mu}^{22})  \; .
\nonumber
\end{eqnarray}
At every link $(x,\mu)$ we have four individual constraints that come from integrating the four phases 
$\phi_{x,\mu}^{(j)}, j = 1,2,4,5$ of the representation (\ref{eq:parametrization}) giving rise to the four 
Kronecker deltas shown in (\ref{linkconstraints}). Here we have 
already taken into account another constraint generated by the $\phi_{x,\mu}^{(3)}$ integral in 
(\ref{eq:partitionsumsu34}) which implements the relation
\begin{equation}
j_{x,\mu}^{21} + j_{x,\mu}^{23} + j_{x,\mu}^{31} + j_{x,\mu}^{33} \; = \; J_{x,\mu}^{12} \; ,
\label{eq:constraintsu31}
\end{equation}
that connects $J_{x,\mu}^{12}$ to the auxiliary currents
$j_{x,\mu}^{ab} =  m_{x,\mu}^{ab} - \overline{m}_{x,\mu}^{ab}$ for the variables 
$m_{x,\mu}^{ab}, \overline{m}_{x,\mu}^{ab}$
introduced in (\ref{eq:binomial}) for the binomial decomposition for the (2,1), (2,3), (3,1) and (3,3) matrix elements. 
To obtain (\ref{linkconstraints}) we have used (\ref{eq:constraintsu31}) to 
replace the combination  $j_{x,\mu}^{21} + j_{x,\mu}^{23} + j_{x,\mu}^{31} + j_{x,\mu}^{33}$ by $J_{x,\mu}^{12}$
in the integrals over $\phi_{x,\mu}^{(j)}, \, j = 1,2,4,5$ in (\ref{eq:partitionsumsu34}). In the final form 
(\ref{linkconstraints}) of $C_G[p]$ we show only the corresponding four constraints, while the 
constraint (\ref{eq:constraintsu31}) is included in the weight factor $W_G[p]$.

The weight factor $W_G[p]$ in  (\ref{eq:partitionsumsu35}) is itself a sum $\sum_{\{l,m,\overline{m}\}}$ 
over configurations of the auxiliary plaquette variables $l_{x,\mu\nu}^{abcd}$
and the link-based auxiliary variables $m_{x,\mu}^{ab}$ and  $\overline{m}_{x,\mu}^{ab}$ 
used for the binomial decomposition in (\ref{eq:binomial}): 
\begin{widetext}
	\begin{align}
	\label{eq:weight}
	& W_G [p] \, = \, 2^{4V} \!\!\! \sum_{\{l,m,\overline{m}\}}  
	\Bigg[ \prod_{x,\mu}  \delta(J_{x,\mu}^{12} - j_{x,\mu}^{21} - j_{x,\mu}^{23} - j_{x,\mu}^{31} - j_{x,\mu}^{33}) \Bigg] \; 
	\Bigg[ \prod_{x,\mu} (-1)^{J_{x,\nu}^{12} + J_{x,\nu}^{23} + J_{x,\nu}^{31} - j_{x,\nu}^{23} - j_{x,\nu}^{31} }\Bigg]
	\\ \nonumber
         & \hspace{61.5mm} \times \! 
	 \Bigg[ \prod_{x,\mu} \prod_{a = 2,3}  \; 
	 \prod_{b=1,3} \!
	\binom{N_{x,\mu}^{ab}}{m_{x,\mu}^{ab}} \! \binom{\overline{N}_{x,\mu}^{ab}}{\overline{m}_{x,\mu}^{ab}} \!
	\Bigg] \,  \Bigg[ \prod_{x,\mu<\nu} \prod_{a,b,c,d} 
	\dfrac{ \left( \beta/2 \right)^{|p_{x,\mu\nu}^{abcd}| + 2\, l_{x,\mu\nu}^{abcd}}}{
	\left(|p_{x,\mu\nu}^{abcd}| + l_{x,\mu\nu}^{abcd}\right)! \; l_{x,\mu\nu}^{abcd}!} \Bigg] 
\\ \nonumber
	& \times \! \Bigg[ \prod_{x,\mu} \B\left(\dfrac{S_{x,\mu}^{11} + S_{x,\mu}^{13} + S_{x,\mu}^{22} + S_{x,\mu}^{32}}{2} + 2\right.\left|\dfrac{S_{x,\mu}^{12} + s_{x,\mu}^{21} +s_{x,\mu}^{23} + s_{x,\mu}^{31} + s_{x,\mu}^{33}}{2} + 1\right)
	\\
	\nonumber
	& \hspace{7.5mm} \times \B\left(\dfrac{S_{x,\mu}^{11} + s^{21}_{x,\nu} + S_{x,\mu}^{23} - s_{x,\mu}^{23} + s_{x,\mu}^{31} + S_{x,\mu}^{33} - s_{x,\mu}^{33}}{2} + 1\right.\left|\dfrac{S_{x,\mu}^{13} + S_{x,\mu}^{21} - s_{x,\mu}^{21} + s_{x,\mu}^{23} + S_{x,\mu}^{31} - s_{x,\mu}^{31} + s_{x,\mu}^{33}}{2} + 1\right)
	\\
	\nonumber
	&\hspace{7.5mm} \times \B\left(\dfrac{s_{x,\mu}^{21} + S^{22}_{x,\nu} + s_{x,\mu}^{23} + S_{x,\mu}^{31} - s_{x,\mu}^{31} + S_{x,\mu}^{33} - s_{x,\mu}^{33}}{2} + 1\right.\left|\dfrac{S_{x,\mu}^{21} - s_{x,\mu}^{21} + S_{x,\mu}^{23} - s_{x,\mu}^{23} + s_{x,\mu}^{31} + S_{x,\mu}^{32} + s_{x,\mu}^{33}}{2} + 1\right)\Bigg].
	\end{align}
\end{widetext}
The configurations of the $m_{x,\mu}^{ab}$ and  $\overline{m}_{x,\mu}^{ab}$ 
are restricted by the Kronecker delta constraints, which implement (\ref{eq:constraintsu31}) at every link $(x,\mu)$. 
In $W_G[p]$ we collect all weights from 
the expansion of the individual Boltzmann factors and the beta functions resulting from the Haar measure integrals. These
weight factors are organized with respect to powers of the inverse gauge coupling $\beta$, i.e., the dual formulation 
in terms of ACC cycle occupation numbers which we develop here 
is a strong coupling expansion. A major advantage of the strong coupling series in terms of 
ACCs is that all weight factors at arbitrary orders of $\beta$ 
are known in closed form: They are given in terms of factorials, binomial coefficients and beta functions (which 
can also be rewritten as fractions of factorials). We stress that in the form 
(\ref{eq:weight}) there is an explicit sign factor. This factor comes from the minus signs in the parametrization
(\ref{eq:parametrization}) used for the SU(3) group elements. This implies that 
for a Monte Carlo simulation of the ACC dual form a strategy for partial resummation needs to be found. 

\begin{figure}
	\vspace*{0.5cm}
	\includegraphics[width=5.4cm]{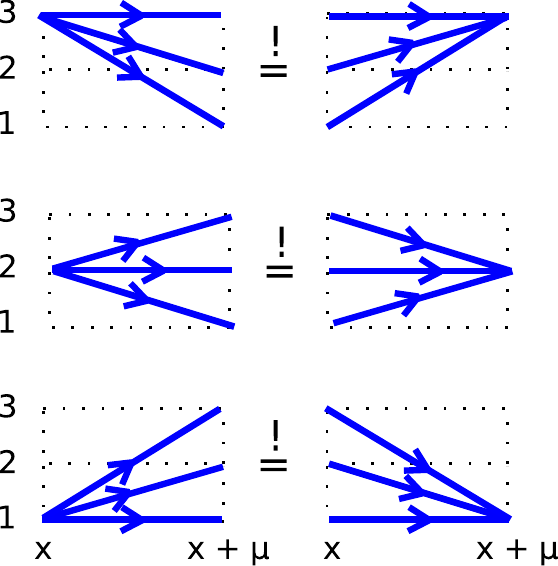}
	\vskip9mm
	\includegraphics[width=8.5cm]{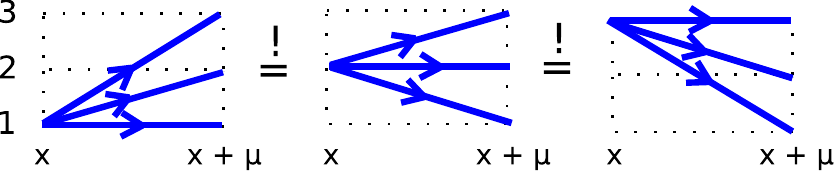}
	\caption{Schematic representation of the three 
	{\sl color conservation constraints} (\ref{eq:constraints_final_1}) -- 		
	(\ref{eq:constraints_final_3}) (top three plots) and the {\sl color exchange constraint} (\ref{eq:constraints_final_4}) 
	(bottom plot). They impose relations between the fluxes $J_{x,\mu}^{ab}$ and 
	admissible configurations of the cycle occupation numbers 
	$p_{x,\mu\nu}^{abcd}$ have to respect these constraints. \label{fig:constraints_su3}}
\end{figure}

Let us now come to the announced discussion of the constraints in (\ref{linkconstraints}). 
Understanding how the SU(3) symmetry of the conventional representation becomes 
manifest in terms of constraints for the dual variables is one of the key points of this paper. 
In (\ref{linkconstraints}) at each link $(x,\mu)$ the fluxes $J_{x,\mu}^{ab}$ are related to each other
by four constraints  implemented by Kronecker deltas. These relations read
\begin{gather}
\label{eq:constraintsu32}
J_{x,\mu}^{12} + J_{x,\mu}^{13} = J_{x,\mu}^{21} + J_{x,\mu}^{31} \; ,\\
\label{eq:constraintsu33}
J_{x,\mu}^{21} + J_{x,\mu}^{23} = J_{x,\mu}^{12} + J_{x,\mu}^{32} \; , \\
\label{eq:constraintsu34}
J_{x,\mu}^{11} + J_{x,\mu}^{12} = J_{x,\mu}^{23} + J_{x,\mu}^{33} \; ,\\
\label{eq:constraintsu35}
J_{x,\mu}^{31} + J_{x,\mu}^{33} = J_{x,\mu}^{12} + J_{x,\mu}^{22} \; .
\end{gather} 
The relations (\ref{eq:constraintsu32}) -- (\ref{eq:constraintsu35}) describe how the SU(3) gauge invariance of the
conventional representation is encoded in constraints for the fluxes $J_{x,\mu}^{ab}$. 
These relations can be combined and 
reorganized in a way that makes the flow between the different color indices $a,b$ more transparent. On both sides of 
(\ref{eq:constraintsu32}) we may add $J_{x,\mu}^{11}$ and on both sides of (\ref{eq:constraintsu33}) 
we add $J_{x,\mu}^{22}$.
Furthermore we can subtract (\ref{eq:constraintsu33}) from (\ref{eq:constraintsu32}) and 
add $J_{x,\mu}^{33}$ on both sides.
This gives the following three relations:
\begin{eqnarray}
\label{eq:constraints_final_1}
J_{x,\mu}^{11} + J_{x,\mu}^{12} + J_{x,\mu}^{13} & = & J_{x,\mu}^{11}  + J_{x,\mu}^{21} + J_{x,\mu}^{31} \; ,\\
\label{eq:constraints_final_2}
J_{x,\mu}^{21} + J_{x,\mu}^{22} + J_{x,\mu}^{23} & = & J_{x,\mu}^{12} + J_{x,\mu}^{22} + J_{x,\mu}^{32} \; ,\\
\label{eq:constraints_final_3}
J_{x,\mu}^{31} + J_{x,\mu}^{32} + J_{x,\mu}^{33} & = & J_{x,\mu}^{13} + J_{x,\mu}^{23} + J_{x,\mu}^{33} \; .
\end{eqnarray}
The first relation (\ref{eq:constraints_final_1}) implies that for all links $(x,\mu)$ the flux out of color 1 at $x$ equals the 
flux into color 1 at $x+\hat{\mu}$. The other two relations imply the same conservation 
law for color 2 and color 3. Thus for all three colors 
$a = 1,2,3$ we have the constraint that along each link the flux out of 
color $a$ has to match the flux into that color $a$. 
Consequently the constraints (\ref{eq:constraints_final_1}) -- (\ref{eq:constraints_final_3}) 
imply that for each color $a$ the flux 
that runs through $a$ is the same at all sites. Thus we refer to 
(\ref{eq:constraints_final_1}) -- (\ref{eq:constraints_final_3}) as
the {\sl ''color conservation constraints''}. These three constraints are 
illustrated in the first three plots of Fig.~\ref{fig:constraints_su3}. 

A second type of constraints among the $J_{x,\mu}^{ab}$ is obtained by adding $J_{x,\mu}^{13}$ on both sides of 
(\ref{eq:constraintsu34}) and adding $J_{x,\mu}^{32}$ on both sides of (\ref{eq:constraintsu35}). 
The right hand sides of the
resulting two equations are then replaced using (\ref{eq:constraints_final_2}) and (\ref{eq:constraints_final_3}), 
and we can summarize the resulting relations as
\begin{equation}
\label{eq:constraints_final_4}
J_{x,\mu}^{11} + J_{x,\mu}^{12} + J_{x,\mu}^{13} = J_{x,\mu}^{21} + J_{x,\mu}^{22} + J_{x,\mu}^{23} 
= J_{x,\mu}^{31} + J_{x,\mu}^{32} + J_{x,\mu}^{33} \, .
\end{equation}
This constraint implies that the flux that flows out of a color $a$ along 
a link $(x,\mu)$ is the same for all three colors $a$. Thus if flux
is exchanged between the colors along some link, the exchanged flux has to be the same for all three colors. We refer to 
(\ref{eq:constraints_final_4}) as {\sl ''color exchange constraints''}.

The relations (\ref{eq:constraints_final_1}) -- (\ref{eq:constraints_final_4}) enforce constraints among the fluxes 
$J_{x,\mu}^{ab}$ that have to be obeyed at every link $(x,\mu)$. The fluxes $J_{x,\mu}^{ab}$ defined in (\ref{eq:Jfluxes}) 
depend only on the cycle occupation numbers $p_{x,\mu\nu}^{abcd} \in \mathbb{Z}$ 
and thus only this set of dual variables is subject to the constraints -- the auxiliary plaquette variables 
$l_{x,\mu\nu}^{abcd} \in \mathbb{N}_0$ are unconstrained. Thus the
three color conservation constraints (\ref{eq:constraints_final_1}) -- (\ref{eq:constraints_final_3}) and the 
color exchange constraint (\ref{eq:constraints_final_4}) implement the original SU(3) symmetry as a set of
constraints that govern the flux of the colors in the dual form of the theory. Identifying these constraints is one of the key 
goals of this paper. 

We stress at this point that the form (\ref{eq:constraints_final_1}) -- (\ref{eq:constraints_final_4}) 
of the constraints is over-complete,
since it contains six relations constructed out of the original four relations  
(\ref{eq:constraintsu32}) -- (\ref{eq:constraintsu35}). However,
the over-complete final form (\ref{eq:constraints_final_1}) -- (\ref{eq:constraints_final_4}) 
where the constraints are not all independent 
better illustrates the complete symmetry among the three colors. 

The constraints imply matching conditions for the color flux along 
the links of the lattice, where originally this flux comes from 
non-zero cycle occupation numbers $p_{x,\mu\nu}^{abcd}$. Thus at 
links where plaquettes touch, the corresponding cycle 
occupation numbers have to be matched such that constraints are obeyed. 

For this matching two cases can be distinguished:
The trivial case where cycle occupation numbers that sit on the same 
plaquette together obey the constraints. Obviously this is only
a local contribution and does not contribute to the long range physics. 

Relevant for the long range physics and thus the continuum limit are the contributions where neighboring 
plaquettes share a link such that the cycle occupation numbers at 
neighboring plaquettes are correlated by the constraints. This 
gives rise to generalized surfaces which we refer to as {\sl ''worldsheets''}. 
To see the worldsheet nature of non-local
admissible configurations we start with setting a single cycle occupation number to $p_{x,\mu\nu}^{abcd} = 1$. 
Clearly this violates the constraints along the four links of the plaquette $(x,\mu\nu)$. 
If one now takes a neighboring plaquette and selects the corresponding 
cycle occupation number such that at the joint link the constraints are obeyed, then 
we have a surface consisting of two plaquettes and 
the contour of links with violated constraints contains six links. One can keep attaching 
plaquettes with suitably chosen plaquette 
occupation numbers to grow the 2-D surface further and the constraints will always be 
violated along the boundary of that surface. 
Thus the only way to obtain a non-local configuration of non-trivial cycle occupation 
numbers, such that all constraints are obeyed, 
is to create a closed surface. Thus the constraints lead to a structure of closed 
worldsheets for admissible configurations of cycle 
occupation numbers. And since the ACCs are already abelian, 
the corresponding cycle occupation numbers are additive and the 
worldsheet picture also holds for cycle occupation numbers with $|p_{x,\mu\nu}^{abcd}| > 1$.
   
Before we come to generalizing the ACC approach to including also fermions, let us briefly 
summarize the dual worldsheet representation 
we have constructed for pure SU(3) lattice gauge theory.  
The partition function of pure SU(3) lattice gauge theory is exactly rewritten 
as a sum over configurations of the cycle occupation numbers $
p_{x,\mu\nu}^{abcd} \in \mathbb{Z}$. At each link $(x,\mu)$ the 
fluxes $J_{x,\mu}^{ab}$ defined in (\ref{eq:Jfluxes}) collect the flux 
of the cycle occupation numbers that connect color $a$ at $x$ to color $b$ at $x + \hat{\mu}$. 
These fluxes are subject to the {\sl color conservation constraints} 
(\ref{eq:constraints_final_1}) -- (\ref{eq:constraints_final_3}) and the 
{\sl color exchange constraint} (\ref{eq:constraints_final_4}). These constraints restrict the admissible configurations of the 
cycle occupation numbers and the long range contributions have the interpretation of worldsheets.

The configurations of the cycle occupation numbers come with weight factors $W[p]$ that are themselves sums 
$\sum_{\{l,m,\overline{m}\}}$ over configurations of the auxiliary plaquette variables $l_{x,\mu\nu}^{abcd}$
and the link-based auxiliary variables $m_{x,\mu}^{ab}$ and  $\overline{m}_{x,\mu}^{ab}$. In these sums the 
constraint (\ref{eq:constraintsu31}) restricts the configurations of the $m_{x,\mu}^{ab}$ and 
$\overline{m}_{x,\mu}^{ab}$ by connecting their differences $j_{x,\mu}^{ab} = m_{x,\mu}^{ab} - \overline{m}_{x,\mu}^{ab}$ 
to $J_{x,\mu}^{12}$. The contributions to $W_G[p]$ collect 
all terms from expanding the Boltzmann factors, as well
as combinatorial factors and the beta functions coming from the Haar measure integration. 
Note that these contributions come with 
signs going back to the signs in the parameterization of the SU(3) matrices (\ref{eq:parametrization}).

All terms in the dual representation are organized with respect to powers of $\beta$, such that the dual 
form (\ref{eq:partitionsumsu35}), (\ref{linkconstraints}), (\ref{eq:weight}) 
constitutes a strong coupling series. We stress again that in this form of the 
strong coupling series all expansion coefficients are known in closed form. 
In the following section we introduce the matter 
fields and show that the ACC approach can be generalized further to obtain a dual form of full QCD.

\section{QCD at Strong Coupling\label{QCDatSC}}

The next step towards a full dual worldline/worldsheet representation of QCD 
is the generalization of the ACC approach to matter fields. 
In order to simplify the presentation we start with an intermediate step where we consider the strong coupling limit. 
In this limit we have $\beta = 0$, i.e., the gauge action is absent. Note that in the strong coupling regime a continuum limit 
cannot be performed. Nevertheless, the strong coupling limit of QCD shares some non-perturbative 
properties with full QCD, such that it is an interesting toy model per se.

For the discussion of the structure of dual worldline representations, which is the main goal of this paper, 
also strong coupling QCD is an interesting theory: Integrating out the SU(3) link variables will again lead to 
constraints for the color fluxes along the links, but in strong coupling QCD these fluxes
are generated by the fermions, instead of the cycle occupation numbers of pure gauge theory.  We will see 
that structurally the constraints are the same, but for strong coupling fermion loops the constraints
are simpler in their interpretation because of the additional restrictions from the Pauli principle.

For simplicity we will consider the derivation for a theory with only one flavor of 
staggered quarks, but stress that the generalization to an arbitrary number of flavors is trivial. 
The fermionic partition function in a background of gauge links is given by
\begin{equation}
	Z_{F}[U] = \int \! \!D \big[\,\overline{\psi},\psi\big] \; e^{-S_{F}[U,\psi,\overline{\psi}]} \, ,
	\label{eq:partition_fermions}
\end{equation}
where $\psi_{x}$ and $\overline{\psi}_{x}$ are 3-component vectors of  Grassmann numbers 
assigned to the sites $x$ of our four-dimensional lattice. They obey anti-periodic boundary 
conditions in the Euclidean time direction ($\nu = 4$) and periodic boundary conditions in space. 
The integration measure in (\ref{eq:partition_fermions}) is a product over 
Grassmann measures 
$\int \! D\big[ \,\overline{\psi},\psi \big] = \prod_{x}\prod_{a = 1}^{3} \int d\overline{\psi}^{a}_{x} d\psi_{x}^{a}$. 
We work with the staggered fermion action given by
\begin{widetext}
	\begin{align}
	\label{eq:fermion_action}
	S_{F}\big[U,\psi,\overline{\psi}\,\big] &= \sum_{x}\biggl[ m \overline{\psi}_{x} \psi_{x} 
	+ \sum_{\nu} \dfrac{\eta_{x,\nu}}{2} \Big( \overline{\psi}_x U_{x,\nu} \psi_{x + \hat{\nu}}\, e^{\mu \delta_{\nu,4}} -\, \overline{\psi}_{x + \hat{\nu}} U_{x,\nu}^{\dagger} \psi_{x}\, e^{-\mu \delta_{\nu,4}} \Big) \biggr]\\
	&= \sum_{x}\biggl[ m \sum_{a = 1}^{3} \overline{\psi}_{x}^{a} \psi_{x}^{a} 
	+ \sum_{\nu} \dfrac{\eta_{x,\nu}}{2} \sum_{a, b = 1}^{3} \left( \overline{\psi}_{x}^{a} U_{x,\nu}^{ab} \psi_{x + \hat{\nu}}^{b}\, e^{\mu \delta_{\nu,4}} -\, \overline{\psi}_{x + \hat{\nu}}^{b} U_{x,\nu}^{ab \, \star} \psi_{x}^{a}\, e^{-\mu \delta_{\nu,4}} \right) \biggr] \, ,
	\nonumber
	\end{align}
\end{widetext}
where $\eta_{x,1} = 1$, $\eta_{x,2} = (-1)^{x_1}$, $\eta_{x,3} = (-1)^{x_1+x_2}$ and $\eta_{x,4} = (-1)^{x_1+x_2+x_3}$ 
are the staggered sign factors. In (\ref{eq:fermion_action}) we also 
introduce a chemical potential $\mu$, which gives a different weight to 
forward and backward hopping in the euclidean time direction. 
The chemical potential will later be useful to identify the particle number in the dual representation in terms of worldlines. 
In the strong coupling limit we are considering in this section, the full partition function is obtained by 
integrating the fermionic partition 
sum with the product Haar measure of the previous section, i.e.,  $Z = \int \! D[U] \, Z_{F}[U]$.

In the first line of (\ref{eq:fermion_action}) we used matrix-vector notation for color, 
while in the second line the sums over color indices were written explicitly. Also for the 
theory with fermions this decomposition of the action is the crucial step 
towards the dualization. It allows one to completely factorize the 
Boltzmann weight such that every term in the last line of 
(\ref{eq:fermion_action}) is a single bilinear in the Grassmann variables 
and thus all terms commute with each other.

Using this decomposition of the fermion action we write the partition function as
\begin{widetext}
	\begin{align}
	\label{eq:partition_fermion2}
		Z_{F} &= \int \!\! D\big[\,\overline{\psi},\psi\big] \!
		\prod_{x} \prod_{a = 1}^{3} e^{- m \overline{\psi}_{x}^{a} \psi_{x}^{a}} \prod_{x, \nu} \prod_{a, b = 1}^{3} e^{- \frac{\eta_{x,\nu}}{2} \overline{\psi}_{x}^{a} U_{x,\nu}^{ab} \psi_{x + \hat{\nu}}^{b} e^{\mu \delta_{\nu,4}}} 
		e^{\frac{\eta_{x,\nu}}{2} \overline{\psi}_{x + \hat{\nu}}^{b} U_{x,\nu}^{ab \, \star} \psi_{x}^{a} e^{-\mu \delta_{\nu,4}}} \\
		\nonumber
		&= \int \!\! D\big[\,\overline{\psi},\psi\big] \! \prod_{x} \prod_{a = 1}^{3} 
		\sum_{s^a_{x} = 0}^{1}\!\! \left( m \psi_{x}^{a} \overline{\psi}_{x}^{a} \right)^{\!s_{x}^{a}} 
		\prod_{x, \nu} \prod_{a, b = 1}^{3} \sum_{\!k_{x,\nu}^{ab} = 0}^{1} 
		\!\!\!\left(\!\frac{-\eta_{x,\nu}}{2}\, \overline{\psi}_{x}^{a} U_{x,\nu}^{ab} 
		\psi_{x + \hat{\nu}}^{b}\, e^{\mu \delta_{\nu,4}} \!\!\right)^{\!k_{x,\nu}^{ab}}  \!\!\!
		\sum_{\overline{k}_{x,\nu}^{ab} = 0}^{1} \!\!\!\! \left( \frac{\eta_{x,\nu}}{2}\, 
		\overline{\psi}_{x + \hat{\nu}}^{b} U_{x,\nu}^{ab \, \star} \psi_{x}^{a}\, 
		e^{-\mu \delta_{\nu,4}} \! \right)^{\!\overline{k}_{x,\nu}^{ab}} \\
		\nonumber
		&= \, \left(\dfrac{1}{2}\right)^{3V} \!\! \sum_{\{s,k,\overline{k}\}} (2m)^{\sum_{x,a} s_{x}^{a}} \, e^{\mu \sum_{x,ab} [k_{x,\hat{4}}^{ab} - \overline{k}_{x,\hat{4}}^{ab}]} 
		\prod_{x,\nu} \prod_{a, b} (-1)^{k_{x,\nu}^{ab}} (\eta_{x,\nu})^{k_{x,\nu}^{ab} + \overline{k}_{x,\nu}^{ab}}	\left( U_{x,\nu}^{ab} \right)^{k_{x,\nu}^{ab}}
		\left( U_{x,\nu}^{ab \, \star} \right)^{\overline{k}_{x,\nu}^{ab}} \\
		&\qquad \qquad \qquad \qquad \qquad \qquad \qquad \quad \times \int \! D\big[\,\overline{\psi},\psi \big] \prod_{x,a} (\psi_{x}^{a} \overline{\psi}_{x}^{a})^{s_{x}^{a}}
			\prod_{x, \nu} \prod_{a, b} \left( \overline{\psi}_{x}^{a}  \psi_{x + \hat{\nu}}^{b} \right)^{k_{x,\nu}^{ab}}
			\left( \overline{\psi}_{x + \hat{\nu}}^{b}  \psi_{x}^{a} \right)^{\overline{k}_{x,\nu}^{ab}} \, .
		\nonumber
\end{align}
\end{widetext}
In the first line we rewrote the Boltzmann weight as a product 
over local exponential factors, which we then Taylor expanded in 
the second step. Note that here the Taylor series terminate after 
the linear term due to the nilpotency of the Grassmann variables. 
We introduce three types of expansion indices (one for every bilinear of the action) 
that will be the new dual variables for the fermions: 
$s_{x}^{a} = 0, 1$ is the dual variable for expanding the color component $a$ of the mass term contribution at site $x$.
$k_{x,\nu}^{ab} = 0, 1$ generates the forward hop from color $a$ to color $b$ on the link $(x,\nu)$, and 
$\overline{k}_{x,\nu}^{ab} = 0, 1$ the backward hop on the same link. When they assume the non-trivial value 1, 
the three types of variables activate the corresponding fermion bilinears: 
The {\sl ''monomer''} variable $s_{x}^{a}$ activates the mass term component with color $a$. 
The dual variables $k_{x,\nu}^{ab}$ 
and $\overline{k}_{x,\nu}^{ab}$ that live on links
activate the forward and backward nearest neighbor bilinears $\overline{\psi}_{x}^{a}  \psi_{x + \hat{\nu}}^{b}$ and 
$\overline{\psi}_{x + \hat{\nu}}^{b}  \psi_{x}^{a}$ that connect color $a$ and $b$ along the link. We refer to these
terms as {\sl ''abelian color fluxes''}.  

\begin{figure*}
		\includegraphics[width=17cm]{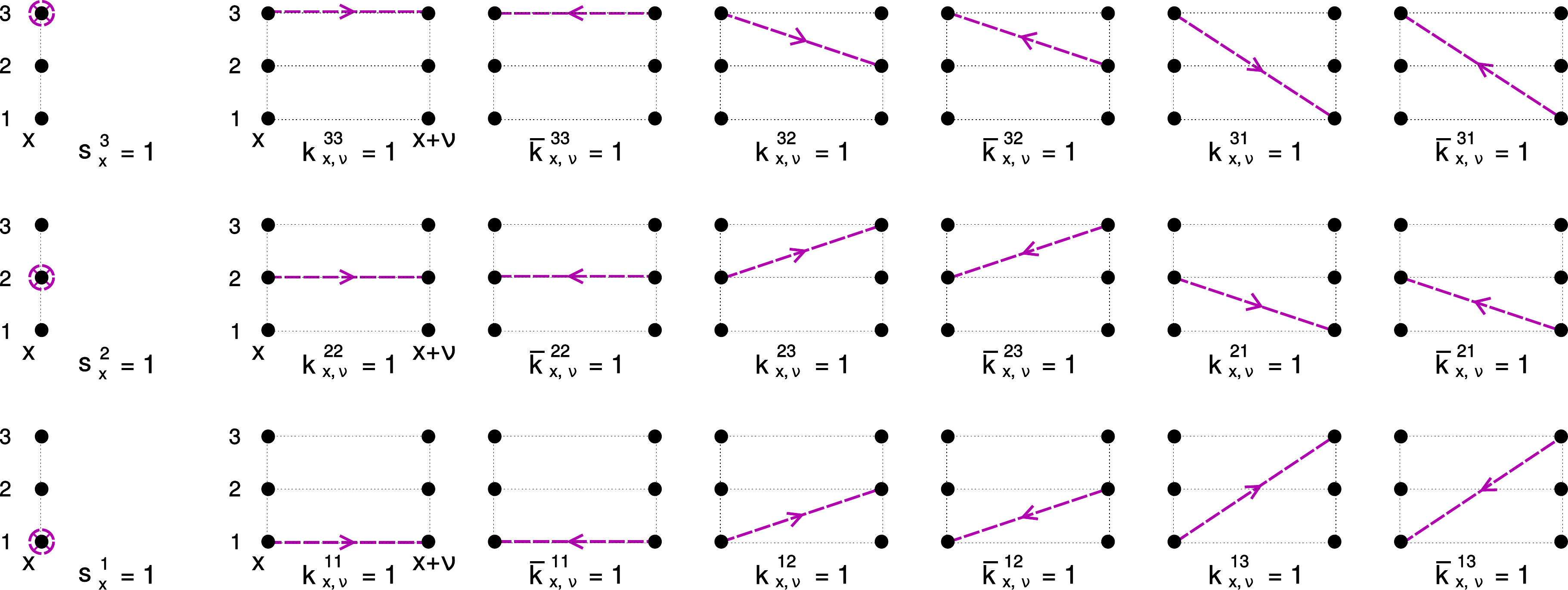}
		\caption{
		Graphical representation of the dual variables for the fermions. As before we use a lattice with three 
		layers to represent the color indices. In the first column of diagrams we show the monomers 
		$s_{x}^{a}$, while the arrows in the other columns represent the dual variables $k_{x,\nu}^{ab}$ and 
		$\overline{k}_{x,\nu}^{ab}$ for the forward and backward hopping, respectively. With these link variables 
		it is possible to build dimers and oriented loops which, together with the monomers, constitute the admissible 
		configurations for fermions. \label{fig:grassmann_dual}}
\end{figure*}

In Fig.~\ref{fig:grassmann_dual} we adapt the graphical representation which we developed for the ACCs now also 
to the fermionic dual variables $s_{x}^{a}$, $k_{x,\nu}^{ab}$ and $\overline{k}_{x,\nu}^{ab}$.
Again we use a lattice with three layers for the three colors. 
The monomers that are activated by $s_{x}^{a}$ sit on a single site 
$x$ and are represented by a circle around the color $a$ they refer to. The {\sl ''link-fluxes''} 
$k_{x,\nu}^{ab}$ and $\overline{k}_{x,\nu}^{ab}$ connect colors $a$ and $b$ 
along the link $(x,\nu)$ and we represent them with
a forward (backward) oriented arrow that connects the color indices $a$ and $b$.

In the last step of Eq.~(\ref{eq:partition_fermion2}) we have already reorganized 
the terms: We have collected an overall factor 
of $(1/2)^{3V}$ and introduced the notation $\sum_{\{s,k,\overline{k}\}}$ for the 
sum over all the possible configurations of the 
fermion dual variables. Finally we write all factors that do not depend on the Grassmann variables in front of the 
Grassmann integral. This Grassmann integral in the last line of (\ref{eq:partition_fermion2}) is either vanishing or $\pm 1$, 
depending on the values of the dual variables $s_{x}^{a}$, $k_{x,\nu}^{ab}$ and $\overline{k}_{x,\nu}^{ab}$. 
In particular, it will be non-vanishing only if each Grassmann variable 
$\psi_{x}^{a} \overline{\psi}_{x}^{a}$ appears exactly once, 
and we refer to this case as \textit{''saturated Grassmann integral''}. 

We may formulate the condition of a saturated Grassmann integral as a constraint for the configurations of the dual 
variables $s_{x}^{a}$, $k_{x,\nu}^{ab}$ and $\overline{k}_{x,\nu}^{ab}$, which can be written in the form
\begin{align}
	\label{eq:fermion_constraint}
	&C_{F}[s,k,\overline{k}] = \\
	\nonumber
	&= \prod_{x,a} \delta \Big( 1 - s_{x}^{a} - \dfrac{1}{2} \sum_{\nu,b} 
	[k_{x,\nu}^{ab} + \overline{k}_{x,\nu}^{ab} + k_{x - \hat{\nu},\nu}^{ba} + \overline{k}_{x - \hat{\nu},\nu}^{ba} ]\Big).
\end{align}
The admissible configurations are known to have a simple structure. 
Here, for the case where we consider a lattice with three 
layers, the admissible configurations are such that in all three layers of the four-dimensional lattice 
each site has to be either 
occupied by a monomer ($s_{x}^{a} = 1$), be the endpoint of a dimer ($k_{x,\nu}^{ab} = \overline{k}_{x,\nu}^{ab} = 1$), 
or be run through by a loop $\mathcal{L}$, 
which is defined as a closed chain of $k_{x,\nu}^{ab} = 1$ and $\overline{k}_{x,\nu}^{ab} = 1$. 

Having discussed the monomer, dimer and loop structure of admissible 
fermion configurations we still need to determine the 
signs of the configurations. Monomers $\psi_{x}^{a}\overline{\psi}_{x}^{a}$ 
are activated by setting $s_{x}^{a} = 1$. It is evident from 
the last equality in Eq.~(\ref{eq:partition_fermion2}) that monomers simply 
come with a factor of $2m$, and the Grassmann 
variables are already in the canonical order we choose for the 
Grassmann integral ($\psi_{x}^{a}$ left of $\overline{\psi}_{x}^{a}$). 
Thus monomers always contribute to admissible configurations with the explicitly positive factor of $2m$. 

Dimers are constructed by setting $k_{x,\nu}^{ab} = \overline{k}_{x,\nu}^{ab} = 1$. 
For the Grassmann integral this corresponds to activating the factor
\begin{equation}
	\label{eq:dimer}
	 \overline{\psi}_{x}^{a} \, \psi_{x + \hat{\nu}}^{b} \, \overline{\psi}_{x + \hat{\nu}}^{b}\, \psi_{x}^{a} \; = \;
	 - \psi_{x}^{a} \, \overline{\psi}_{x}^{a} \, \psi_{x + \hat{\nu}}^{b} \, \overline{\psi}_{x + \hat{\nu}}^{b} \; .
\end{equation}
The minus sign on the right hand-side of (\ref{eq:dimer}) results from the reordering 
of the Grassmann variables into the canonical order. 
However, this minus sign is compensated by the explicit minus sign for the forward hop, 
which in (\ref{eq:partition_fermion2}) is taken into account in the factor 
$(-1)^{k_{x,\nu}^{ab}} = (-1)^{1} = -1$. Also the staggered sign factor contribution is always positive for dimers, 
since $(\eta_{x,\nu})^{k_{x,\nu}^{ab} + \overline{k}_{x,\nu}^{ab}} = (\eta_{x,\nu})^{2} = 1$. 
Finally, also a possible minus sign from the 
anti-periodic temporal boundary conditions for the fermions is irrelevant since 
for a dimer such a sign would appear twice. Thus, dimers always come 
with a positive sign and only loops can generate negative signs.

The overall sign of a loop receives several contributions. Each loop $\mathcal{L}$ picks 
up a minus sign from commuting the Grassmann 
variables into the canonical order. Moreover, each forward hop of the 
loop will contribute with a minus sign. Hence, if $|\mathcal{L}|$ 
denotes the length of the loop $\mathcal{L}$, the sign coming from the 
forward hops is $(-1)^{|\mathcal{L}|/2}$, since half of the hops in a closed loop
are in forward direction. An exception are loops that wind around the compact boundaries, 
but we may restrict ourselves to choosing lattice extents that are a multiple of 4, in which case 
$(-1)^{|\mathcal{L}|/2}$ is correct also for loops that wind. Related to the winding 
of the loops is also the sign that is generated by the anti-periodic temporal boundary conditions. 
For every crossing of the last temporal link an additional factor of $-1$ has to be taken into account, 
such that the sign factor $(-1)^{W_\mathcal{L}}$ emerges, where $W_\mathcal{L}$ is the temporal winding 
number of the loop $\mathcal{L}$.

Finally we have to determine the sign that comes from the staggered sign factors along the links of the loop.
Let us first consider a loop around a single plaquette $(x,\rho \nu)$. Around the plaquette the contribution from the 
staggered signs is given by
\begin{equation}
	\label{eq:staggered_plaquette}
	\eta_{x,\rho} \, \eta_{x + \hat{\rho},\nu}\, \eta_{x + \hat{\nu},\rho} \, \eta_{x,\nu} \; = \; -1 \, ,
\end{equation}
and moreover this factor of $-1$ is independent of the position and orientation of the plaquette. 
If we then consider two adjacent plaquettes, the staggered sign on the common link will cancel 
out from the product of staggered factors because it gets squared. Thus the sign of two adjacent plaquettes 
is also the sign from the staggered factors for the loop along the boundary of the two plaquettes. 
This mechanism can be iterated to construct a loop of any shape, and the 
sign coming from the staggered factor can be expressed as $(-1)^{P_\mathcal{L}}$, where 
$P_\mathcal{L}$ is the number of plaquettes in the surface 
bounded by the loop $\mathcal{L}$. Due to the fact that we have three layers of colors, 
the admissible configurations may also contain loops 
that wind around the same contour up to three times (see Fig.~\ref{fig:loop} for a 
simple example of such a loop). For these cases 
we need a multiply covered surface (e.g., a surface that is covered 3-times 
for the example in the bottom plot of Fig.~\ref{fig:loop}) and the 
total number $P_\mathcal{L}$ of plaquettes in the surface spanned by the loop is 
understood in the sense that it also takes into account 
multiple coverings. We finally remark that the surface that has a loop $\mathcal{L}$ 
as its boundary is not unique, but it is easy to see that different 
surfaces with the same boundary differ by an even number of plaquettes, 
such that the sign factor $(-1)^{P_\mathcal{L}}$ remains unchanged.

We can summarize our discussion of the admissible fermion configurations as follows: 
Admissible configurations are those where every site in our 3-layer lattice is either occupied 
by a monomer, is the endpoint of a dimer, or is run through by a loop. 
Monomers ($s_{x}^{a} = 1$) come with a factor of $2m$. Dimers 
($k_{x,\nu}^{ab} = \overline{k}_{x,\nu}^{ab} = 1$) come with a 
factor of 1, but also activate the SU(3) matrix elements along the link, i.e., they
activate the factor $U_{x,\nu}^{ab} U_{x,\nu}^{ab \, \star}$,
that will contribute in the Haar measure integration. Finally loops 
${\cal L}$ come with a sign factor, which, following the discussion above,
is given by 
\begin{equation}
	\label{eq:loop_sign}
	\sign(\mathcal{L}) = (-1)^{1 + |\mathcal{L}|/2 + P_\mathcal{L} + W_\mathcal{L}} \; ,
\end{equation}
where $|\mathcal{L}|$ is the length of the loop $\mathcal{L}$, $P_\mathcal{L}$ is the number of 
plaquettes necessary to cover the surface bounded by the loop 
$\mathcal{L}$, and $W_\mathcal{L}$ is the number of temporal windings of $\mathcal{L}$. 

To obtain the full strong coupling partition sum $Z = \int \! D[U] Z_{F}[U]$, we still have to integrate the fermionic 
partition function $Z_F[U]$ over the product of SU(3) Haar measures.  We find
\begin{align}
	\label{eq:partition}	
	Z& = \sum_{\{s,k,\overline{k}\}} C_{F}\big[s,k,\overline{k}\big] \; W_{F}\big[s,k,\overline{k}\big] 
         \\
	& \qquad \times \int \! D[U]
	\prod_{x, \nu} \prod_{a, b} 	\left( U_{x,\nu}^{ab} \right)^{k_{x,\nu}^{ab}}
	\left( U_{x,\nu}^{ab \, \star} \right)^{\overline{k}_{x,\nu}^{ab}} \, ,
	\nonumber
\end{align}
where we introduced the weight for the fermion configurations $W_{F}[s,k,\overline{k}]$ defined as
\begin{eqnarray}
&& W_{F}\big[s,k,\overline{k}\big] \; = 
\label{fermionweight}
\\
&& \left( \frac{1}{2}\right)^{\!\!3V} \!\! \prod_{\mathcal{L}} \sign(\mathcal{L}) 
\prod_{x} \bigg[ \prod_a (2m)^{s_{x}^{a}} \bigg] \bigg[ \prod_{ab} 
e^{\mu [k_{x,{4}}^{ab} - \overline{k}_{x,{4}}^{ab} ]} \bigg]\, =
\nonumber
\\
&& \left( \frac{1}{2}\right)^{\!\!3V} \!\! \prod_{\mathcal{L}} \sign(\mathcal{L}) \, 
e^{\mu \beta W_\mathcal{L}} \,  \bigg[  \prod_{x,a} (2m)^{s_{x}^{a}} \bigg] \, . 
\nonumber
\end{eqnarray}
In the last step we have simplified the term that couples to the chemical potential $\mu$: The chemical potential 
multiplies the difference $k_{x,{4}}^{ab} - \overline{k}_{x,{4}}^{ab}$ of the temporal forward and backward 
fluxes. For dimers this difference is zero such that they do 
not couple to $\mu$.  Thus only fermion loops contribute to the $\mu$-dependence. The fermion loops are made from 
chains of $k_{x,\nu}^{ab}$ and $\overline{k}_{x,\nu}^{ab}$ where at each site the flux is conserved. Consequently only
loops that wind around the compact time direction can have non-vanishing 
$\sum_x \sum_{ab} ( k_{x,{4}}^{ab} - \overline{k}_{x,{4}}^{ab})$, and it is obvious that for a loop 
$\mathcal{L}$ this sum is given by 
$N_t \, W_\mathcal{L}$, where $N_t$ is the temporal extent of the lattice and $W_\mathcal{L}$ 
is the temporal winding number
of the loop $\mathcal{L}$. Using the fact that the inverse temperature $\beta$ in lattice units is given by $N_t$ we end up 
with the expression for the coupling to $\mu$ given in the last line of (\ref{fermionweight}). 

Comparing the $\mu$-dependence in the last line of (\ref{fermionweight})
with the usual form $e^{\mu \beta {\cal N}}$ 
for the coupling of the chemical potential, where ${\cal N}$ is the net-particle number,
we conclude that the net particle number is given by ${\cal N} = \sum_\mathcal{L} W_\mathcal{L}$. 
Thus we find a nice geometrical interpretation of the net particle number ${\cal N}$ in the worldline formulation: 
${\cal N}$ is given by the total temporal net-winding number of all fermion loops. 

We stress that this identification of the net particle number as a 
topological quantity, i.e., the total temporal net winding number
of the loops, is quite different from the manifestation of the particle number in the conventional representation: There
the net particle number is given by the discretized integral over the zero component of the conserved vector current,
clearly a quantity that is challenging to determine and usually not an integer. In the worldline representation, on the 
other hand, the temporal winding number is a simple quantity and it is very easy to define the canonical ensemble 
by the class of configurations with a fixed temporal net winding number of the fermion loops. 
We consider this to be one of the most beautiful geometrical
aspects of the worldline formulation of QCD. Furthermore, in a toy model it was demonstrated recently that the
simple form of the net particle number can be used to implement worldline simulations of the canonical ensemble
\cite{Orasch:2017niz,Giuliani:2017qeo}.   

Having completed the discussion of the fermionic part let us now continue with the remaining gauge integration.
The integral over the gauge fields in the last line of (\ref{eq:partition}) 
can be done in the same way as the corresponding integrals in the 
pure gauge theory case discussed in the previous section. We insert 
the path integral measure $D[U]$ and the explicit parametrization 
(\ref{eq:parametrization}) for the matrix elements $U_{x,\nu}^{ab}$. 
For those matrix elements that are sums of complex numbers 
we use again the binomial representation (\ref{eq:binomial}). 
However, since $k_{x,\nu}^{ab}, \overline{k}_{x,\nu}^{ab}, m_{x,\nu}^{ab}, \overline{m}_{x,\nu}^{ab} \in \{0,1\}$ 
all binomial factors $\binom{k^{ab}_{x,\mu}}{m^{ab}_{x,\mu}}$ and 
$\binom{\overline{k}^{ab}_{x,\mu}}{\overline{m}^{ab}_{x,\mu}}$ 
are  equal to 1 and we can drop them here. Hence, 
for the partition function of strong coupling QCD we obtain
\begin{equation}
	Z = \!\! \sum_{\{s,k,\overline{k}\}}  C_{F}[s,k,\overline{k}] \; W_F[s,k,\overline{k}] \; 
	C_{G}[k,\overline{k}] \; W_G [k,\overline{k}]  \, .
	\label{eq:partitionsc}
\end{equation}
The gauge field integration in (\ref{eq:partition}) has generated a link based gauge constraint 
$C_{G}[k,\overline{k}]$ and a gauge field weight factor $W_G [k,\overline{k}]$. 
To represent the constraints and the weight factor
in a transparent way, we introduce combinations of the dual variables $k_{x,\nu}^{ab}, \overline{k}_{x,\nu}^{ab}$ and the 
auxiliary variables $m_{x,\nu}^{ab}, \overline{m}_{x,\nu}^{ab}$ for $(a,b) = (2,1), \, (2,3), \, (3,1), \, (3,3)$
as follows:
\begin{eqnarray}
	\label{eq:kshorthand}
	K_{x,\nu}^{ab} & = & k_{x,\nu}^{ab} - \overline{k}_{x,\nu}^{ab} \; , \qquad  
	P_{x,\nu}^{ab} = k_{x,\nu}^{ab} + \overline{k}_{x,\nu}^{ab} \; ,
	\\
	j_{x,\nu}^{ab} \,  & = & 
	m_{x,\nu}^{ab} - \overline{m}_{x,\nu}^{ab} \; , \quad \; \,
	s_{x,\nu}^{ab} = m_{x,\nu}^{ab} + \overline{m}_{x,\nu}^{ab} \, .
	\nonumber
\end{eqnarray}
Again constraints are generated by the integration over the phases $\phi_{x,\nu}^{(j)}$ of the representation
(\ref{eq:parametrization}), and as in the pure gauge case we
can organize them such that four of them give rise to relations among the fluxes $K_{x,\nu}^{ab}$. These 
gauge field constraints are denoted by $C_{G}[k,\overline{k}]$ in (\ref{eq:partitionsc}) and are explicitly given by   
\begin{eqnarray}
\label{eq:constraintsSC}
C_{G} [k,\overline{k}]  & = &  \prod_{x,\nu}  \delta(K_{x,\mu}^{12} + K_{x,\nu}^{13} - K_{x,\nu}^{21} - K_{x,\nu}^{31}) 
\\
&& \hspace{1.2mm} \times  \, 
\delta(K_{x,\nu}^{21} + K_{x,\nu}^{23} - K_{x,\nu}^{12} - K_{x,\nu}^{32}) 
\nonumber \\ 
&& \hspace{1.2mm} \times \, 
\delta(K_{x,\nu}^{11} + K_{x,\nu}^{12} - K_{x,\nu}^{23} - K_{x,\nu}^{33})
\nonumber \\ 
&& \hspace{1.2mm} \times \, 
\delta(K_{x,\nu}^{31} + K_{x,\nu}^{33} - K_{x,\nu}^{12} - K_{x,\nu}^{22})  \; .
\nonumber
\end{eqnarray}
The weight factor $W_G [k,\overline{k}]$ is given as a sum over configurations of the 
$m_{x,\nu}^{ab}, \overline{m}_{x,\nu}^{ab}$
and contains another constraint, 
\begin{equation}
K_{x,\nu}^{12} \; = \;  j_{x,\nu}^{21} + j_{x,\nu}^{23} + j_{x,\nu}^{31} + j_{x,\nu}^{33} \; ,
\label{constraint_Kjjjj}
\end{equation}
which comes from integrating over $\phi_{x,\nu}^{(3)}$ and connects the 
sum of auxiliary variables $j_{x,\nu}^{ab}$ to $K_{x,\nu}^{12}$. Explicitly the 
weight factor is given by
\begin{widetext}
	\begin{align}
	\label{eq:weightHSC}
	W_{G} &[k,\overline{k}] \, = \,  2^{4V} \!\!\! \sum_{\{m,\overline{m}\}}
	 \Bigg[ \prod_{x,\nu} 
	 \delta \left(K_{x,\nu}^{12} - j_{x,\nu}^{21} - j_{x,\nu}^{23} - j_{x,\nu}^{31} - j_{x,\nu}^{33} \right) \Bigg]
	 \Bigg[ \prod_{x,\nu} (-1)^{K_{x,\nu}^{12} + K_{x,\nu}^{23} + K_{x,\nu}^{31} - j_{x,\nu}^{23} - j_{x,\nu}^{31} } \Bigg]
	\\
	&\hspace{-6.5mm}\times \! \Bigg[		
	 \prod_{x,\nu}  \B\! \left(\dfrac{P_{x,\nu}^{11} + P_{x,\nu}^{13} + P_{x,\nu}^{22} + P_{x,\nu}^{32}}{2} + 2
	 \! \right.\left|
	 \dfrac{P_{x,\nu}^{12} + s_{x,\nu}^{21} +s_{x,\nu}^{23} + s_{x,\nu}^{31} + s_{x,\nu}^{33}}{2} + 1\!\right)
	\nonumber \\
	&\; \times \B\! \left(\dfrac{P_{x,\nu}^{11} + s^{21}_{x,\nu} + P_{x,\nu}^{23} - 
	s_{x,\nu}^{23} + s_{x,\nu}^{31} + P_{x,\nu}^{33} - s_{x,\nu}^{33}}{2} + 1
	\! \right.\left|\dfrac{P_{x,\nu}^{13} + P_{x,\nu}^{21} - s_{x,\nu}^{21} + s_{x,\nu}^{23} + 
	P_{x,\nu}^{31} - s_{x,\nu}^{31} + s_{x,\nu}^{33}}{2} + 1\!\right)
	\nonumber \\
	&\; \times \B\! \left(\dfrac{s_{x,\nu}^{21} + P^{22}_{x,\nu} + s_{x,\nu}^{23} + P_{x,\nu}^{31} - 
	s_{x,\nu}^{31} + P_{x,\nu}^{33} - s_{x,\nu}^{33}}{2} + 1
	\! \right.\left|
	\dfrac{P_{x,\nu}^{21} - s_{x,\nu}^{21} + P_{x,\nu}^{23} - s_{x,\nu}^{23} + 
	s_{x,\nu}^{31} + P_{x,\nu}^{32} + s_{x,\nu}^{33}}{2} + 1\! \right)\! \Bigg] ,
	\nonumber
\end{align}
\end{widetext}
where we have defined
\begin{equation}
\sum_{\{m, \overline{m}\}} =  \; \prod_{x,\nu} \; \prod_{a = 2,3}  \; \prod_{b=1,3} \; 
\sum_{m_{x,\nu}^{ab} = 0}^{k_{x,\nu}^{ab}} \; \sum_{\overline{m}_{x,\nu}^{ab} = 0}^{\overline{k}_{x,\nu}^{ab}}  \; .
\end{equation}
Having completed the derivation of the dual representation for strong coupling QCD collected in
Eqs.~(\ref{eq:partitionsc}), (\ref{eq:constraintsSC}) and (\ref{eq:weightHSC}), 
it is highly instructive to discuss the structural similarity with the dual representation of the 
pure gauge theory case in Eqs.~(\ref{eq:partitionsumsu35}), (\ref{linkconstraints}) and (\ref{eq:weight}).

In both, the pure gauge theory and the strong coupling QCD cases, we have color flux that 
lives on the links of the lattice and connects the 3 different color labels on both ends of the link. In 
pure gauge theory this flux is generated by the plaquette-based cycle occupation numbers $p_{x,\mu \nu}^{abcd}$, 
which contribute to the fluxes on all four links of the plaquette. Consequently the pure gauge theory partition sum 
(\ref{eq:partitionsumsu35}) is a sum over all configurations of the cycle 
occupation numbers. In the strong coupling QCD case 
the color flux on the links is generated by fermion loops. These fermion loops are described by the dual fermion variables 
$k_{x,\mu}^{ab}$ and $\overline{k}_{x,\mu}^{ab}$, which together with the monomer variables $s^{a}_x$ have to obey
the fermion constraints $C_F[s,k,\overline{k}]$ in (\ref{eq:fermion_constraint}). 
The fermion constraints force the variables $k_{x,\mu}^{ab}$ and 
$\overline{k}_{x,\mu}^{ab}$ to form closed loops of color flux. These fluxes around closed loops may be viewed as 
generalizations of the fluxes in pure gauge theory which generated by non-zero cycle occupation numbers 
$p_{x,\mu \nu}^{abcd}$ and thus are only around single plaquettes.

Having understood that in both cases we deal with link based fluxes around closed loops (plaquettes or general loops),
we can now compare the gauge field weight factors and the constraints. A comparison of the weight factor $W_G[p]$  
of the pure gauge theory in Eq.~(\ref{eq:weight}) and the weight factor $W_G[k,\overline{k}]$ for strong coupling QCD 
in (\ref{eq:weightHSC}) shows their structural similarity. Both are sums over configurations of the auxiliary variables 
$m_{x,\nu}^{ab}$ and $\overline{m}_{x,\nu}^{ab}$ needed for the binomial decomposition. In both cases the same
auxiliary constraint connects the configurations of these via the combination 
$j_{x,\nu}^{ab} = m_{x,\nu}^{ab} - \overline{m}_{x,\nu}^{ab}$ to the $K_{x,\nu}^{12}$ color flux at every link. 
Furthermore, the same sign factors appear in the summands of both weight factors. Due to the Pauli principle, the fluxes 
in the strong coupling case are restricted to the values 0,1 and $-1$, 
such that all binomial coefficients are equal to 1, while in the 
pure gauge theory weight $W_G[p]$ in Eq.~(\ref{eq:weight}) the binomial coefficients can have non-trivial values. 
Furthermore, in the pure gauge 
theory weight $W_G[p]$ we have plaquette based weight factors from the expansion of the gauge action which also 
depend on the auxiliary plaquette variables $l_{x,\mu \nu}^{abcd}$. Clearly these terms are absent in strong coupling 
QCD where we have no gauge action.

However, the weight factors that come from the Haar measure integration, and thus (together with the constraints)
are responsible for implementing the SU(3) symmetry in the worldline representation, are identical in the two cases: 
They are given as the product of the three beta functions that appear in (\ref{eq:weight}) and in (\ref{eq:weightHSC}) 
and come from integrating the three angles $\theta_{x,\nu}^{(j)}$ with the corresponding Haar measure contributions. 
Obviously, in (\ref{eq:weight}) and in (\ref{eq:weightHSC}) these weights also couple to the same color flux components.

Also the second ingredient that is necessary to implement the SU(3) symmetry of the conventional representation, i.e.,
the constraints, are the same for the pure gauge theory and strong coupling QCD. 
From (\ref{eq:constraintsSC}) we read off the relations
\begin{gather}
\label{eq:constraintsSC2A}
K_{x,\nu}^{12} + K_{x,\nu}^{13} = K_{x,\nu}^{21} + K_{x,\nu}^{31} \; ,\\
K_{x,\nu}^{21} + K_{x,\nu}^{23} = K_{x,\nu}^{12} + K_{x,\nu}^{32} \;  \\
K_{x,\nu}^{11} + K_{x,\nu}^{12} = K_{x,\nu}^{23} + K_{x,\nu}^{33} \; ,\\
K_{x,\nu}^{31} + K_{x,\nu}^{33} = K_{x,\nu}^{12} + K_{x,\nu}^{22} \; ,
\label{eq:constraintsSC2B}
\end{gather} 
which are structurally identical to those for the fluxes $J_{x,\nu}^{ab}$ of pure gauge theory in 
(\ref{eq:constraintsu32}) -- (\ref{eq:constraintsu35}). Thus we can 
recombine them in the same way and bring them to the form of 
(\ref{eq:constraints_final_1}) -- (\ref{eq:constraints_final_4}), giving rise to the same geometrical
interpretation which we illustrated in Fig.~\ref{fig:constraints_su3}. 

The structural similarity for the constraints and the weights we have discussed constitutes the essence of the 
dual worldline/worldsheet representation for systems with SU(3) gauge fields. Other aspects, such as implications
of the constraints for the matter field worldlines are specific for the type of matter the gauge links couple to. Let us now 
address this aspect in more detail for the case of strong coupling QCD, and discuss the structure of strong 
coupling fermion loops.

\section{Strong coupling Baryon fluxes}

For the case of strong coupling QCD, only the dual fermion variables $k_{x,\nu}^{ab}$ and 
$\overline{k}_{x,\nu}^{ab}$ generate color flux. Since these variables can only be 0 or 1, the
corresponding color flux variables $K_{x,\nu}^{ab} = k_{x,\nu}^{ab} - \overline{k}_{x,\nu}^{ab}$ 
that enter the constraints (\ref{eq:constraintsSC2A}) --  (\ref{eq:constraintsSC2B}) are restricted 
to the values $-1,0,1$, where $K_{x,\nu}^{ab}  = +1$ corresponds to a unit flux from color $a$ at $x$ 
into color $b$ at $x +\hat{\nu}$ and $K_{x,\nu}^{ab}  = -1$ to the corresponding flux in the opposite direction. 
Thus we have only a small number of possible color flux configurations on a link $(x,\nu)$ which are 
further restricted by the constraints in Eqs.~(\ref{eq:constraintsSC2A}) -- (\ref{eq:constraintsSC2B}).

Also the auxiliary variables $m_{x,\nu}^{ab}$ and $\overline{m}_{x,\nu}^{ab}$ for the binomial decomposition 
which we sum over in the gauge field weight $W_G[k,\overline{k}]$ in (\ref{eq:weightHSC}) are highly restricted since 
$0 \leq m_{x,\nu}^{ab} \leq k_{x,\nu}^{ab} \leq 1$ and $0 \leq \overline{m}_{x,\nu}^{ab} \leq \overline{k}_{x,\nu}^{ab} \leq 1$. 
Furthermore, via $j^{ab}_{x,\nu} = m^{ab}_{x,\nu} - \overline{m}^{ab}_{x,\nu}$ they are restricted further
by the constraint
\begin{equation}
 j_{x,\nu}^{21} + j_{x,\nu}^{23} + j_{x,\nu}^{31} + j_{x,\nu}^{33} \; = \; K_{x,\nu}^{12} \; , 
 \label{auxconstraint}
 \end{equation}
 that appears in the gauge field weight $W_G[k,\overline{k}]$ in (\ref{eq:weightHSC}). Since both, the dual fermion 
 variables $k_{x,\nu}^{ab}$, $\overline{k}_{x,\nu}^{ab}$, as well as the auxiliary variables 
 $m_{x,\nu}^{ab}$, $\overline{m}_{x,\nu}^{ab}$ are highly restricted in strong coupling QCD, we can completely 
 list all flux combinations that are admissible at a given link. 
 In addition we can determine the corresponding sign that appears in 
 the weight $W_G[k,\overline{k}]$, which for a link $(x,\nu)$ is given by
 \begin{equation}
	\label{eq:gaugesignSC}
	(-1)^{K_{x,\nu}^{12} + K_{x,\nu}^{23} + K_{x,\nu}^{31} - j_{x,\nu}^{23} - j_{x,\nu}^{31} } .
\end{equation}
\begin{figure}
\includegraphics[width=7.5cm]{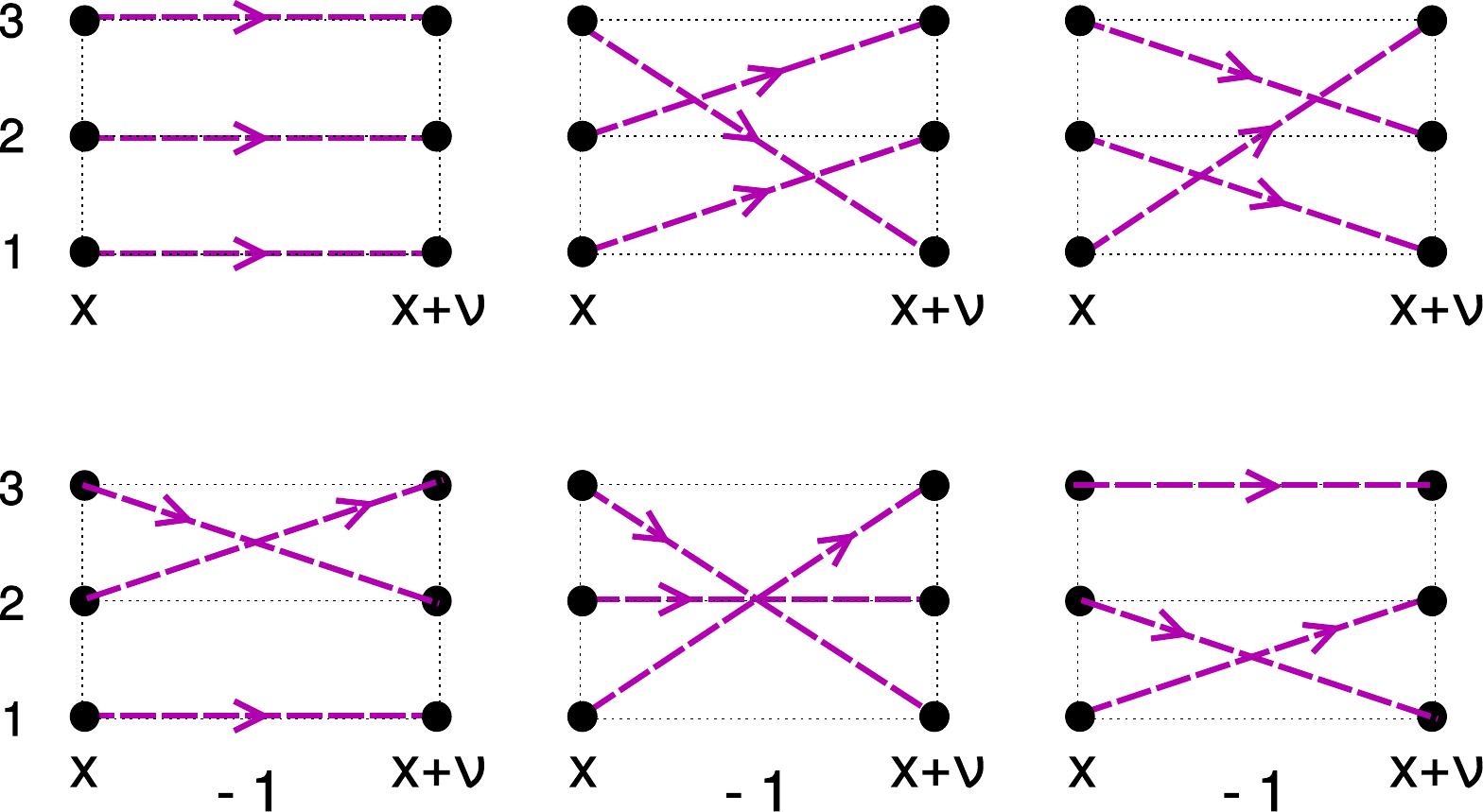}
\caption{Baryon loop elements in the strong coupling limit. Only the six combinations shown here are admissible 
and propagating fluxes in the strong coupling limit. The elements with an odd number of color flux crossings come 
with an explicit minus sign. For the negative direction the same fluxes are admissible and have the same signs.
The corresponding diagrams are obtained by reverting the arrows.
\label{fig:loop_elements}}
\end{figure}

\begin{figure}
	\includegraphics[width=8.5cm]{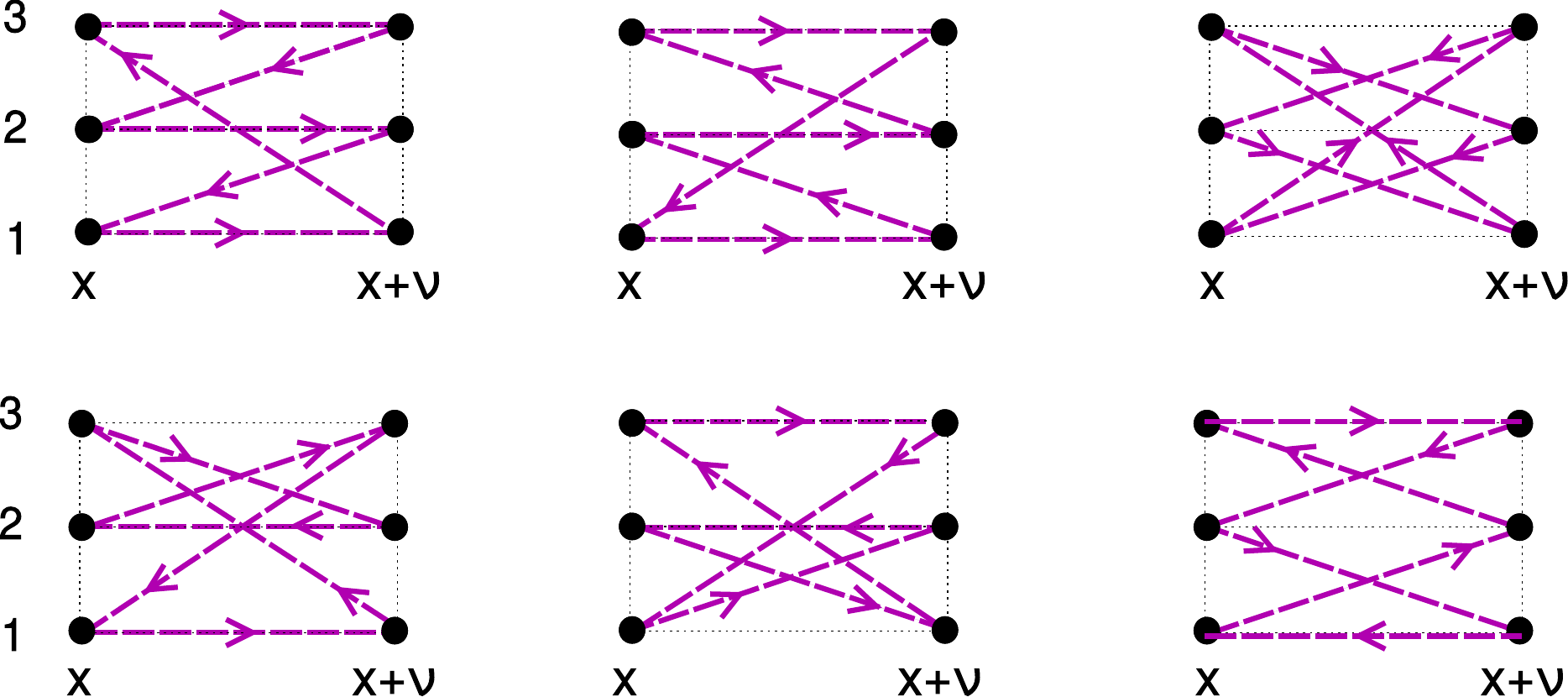}
	\caption{Closed, non-propagating one-link loops at strong coupling.  All of these loops come 
	with a positive weight. Also the opposite orientation is possible, which is obtained by reverting all arrows.
	\label{fig:link_loops}}
\end{figure}

The admissible combinations of the strong coupling fluxes $K_{x,\nu}^{ab}$ at a single link come in two types: 
Three lines of flux that run in the same direction (see Fig.~\ref{fig:loop_elements}), or six lines of flux that
form a closed
loop on a single link (Fig.~\ref{fig:link_loops}). Obviously only the first type allows for long distance propagation 
and we refer to these strong coupling elements as {\sl ''strong coupling baryon fluxes''}. The locally closing ones
are referred to as {\sl ''one link loops''}.

For the discussion of the complete list of strong coupling baryon fluxes we start with solutions of the constraint equations 
(\ref{eq:constraintsSC2A}) --  (\ref{eq:constraintsSC2B}) where we allow only the values $K_{x,\nu}^{ab} = 1,0$,
i.e., we consider forward propagation. In addition to the gauge constraints also 
the fermion constraints have to be obeyed, which 
imply that from a node with fixed space time $x$ and fixed color $a$ only a single forward arrow may origin. 
One finds exactly six solutions, and Fig.~\ref{fig:loop_elements} shows the admissible strong coupling baryon fluxes 
for forward propagation. The six strong coupling baryon fluxes for backward propagation are obtained 
by reverting the arrows, which corresponds to $K_{x,\nu}^{ab} \rightarrow - K_{x,\nu}^{ab}$. 

The signs (\ref{eq:gaugesignSC}) are easy to determine for these six configurations. Let us discuss two examples: 
For the top left example in Fig.~\ref{fig:loop_elements} we have the non-vanishing fluxes 
$K_{x,\nu}^{11} =  K_{x,\nu}^{22} = K_{x,\nu}^{33} = 1$. Since $K_{x,\nu}^{23} =  K_{x,\nu}^{31} = 0$, also  
$j_{x,\nu}^{23}$ and  $j_{x,\nu}^{31}$ must vanish, such that the sign (\ref{eq:gaugesignSC}) is $+1$.

The top center example in Fig.~\ref{fig:loop_elements} has the non-vanishing fluxes 
$K_{x,\nu}^{12} =  K_{x,\nu}^{23} = K_{x,\nu}^{31} = 1$, and these three terms alone give a minus sign in
(\ref{eq:gaugesignSC}). However, since $K_{x,\nu}^{12} = 1$ we must also have one of the $j_{x,\nu}^{21}$,
$j_{x,\nu}^{23}$, $j_{x,\nu}^{31}$, or $j_{x,\nu}^{33}$ to be set to 1, in order to obey the additional constraint 
$(\ref{auxconstraint})$. Since only $K_{x,\nu}^{23}$ and $K_{x,\nu}^{31}$ are non-zero, either 
$j_{x,\nu}^{23}$ or $j_{x,\nu}^{31}$ must be 1, and either choice brings the total sign in (\ref{eq:gaugesignSC})
back to $+1$. 

In a similar way one may analyze the sign for all strong coupling baryon fluxes and one finds the simple result
\begin{equation}
	\label{eq:gaugesignSC2}
	(-1)^{K_{x,\nu}^{12} + K_{x,\nu}^{23} + K_{x,\nu}^{31} - j_{x,\nu}^{23} - j_{x,\nu}^{31} } \; = \;  
	(-1)^{\text{\# crossings of $K$-flux}} \; .
\end{equation} 
In Fig.~\ref{fig:loop_elements} the strong coupling baryon fluxes where this sign is negative are marked 
with $-1$. In addition one may evaluate the weight given by the product of beta functions in (\ref{eq:weightHSC}) 
and a simple calculation shows that this weight has the value of $1/12$ for all 6 strong coupling baryon fluxes 
shown in Fig.~\ref{fig:loop_elements}.

The second class of solutions of (\ref{eq:constraintsSC2A}) --  (\ref{eq:constraintsSC2B}), i.e., the 
one link loops are obtained by now allowing all values $K_{x,\nu}^{ab} = -1,0,+1$ and enforcing the fermion
constraints, such that each node is run through by a loop. The corresponding solutions are depicted in  
Fig.~\ref{fig:link_loops} and it is easy to see that they are obtained by combining one of the forward baryon fluxes from
Fig.~\ref{fig:loop_elements} with a matching backward baryon flux such that the fermion constraints are obeyed. One finds
that only the fluxes with the same sign in Fig.~\ref{fig:loop_elements} can be combined among each other,
such that the total sign from (\ref{eq:gaugesignSC}) is always +1. According to (\ref{eq:loop_sign}) the emerging 
loops also have a positive fermion loop sign: There is an an overall minus sign and a factor 
$(-1)^3$ for the three forward hops. Thus one link loops always come with a positive weight which is given by 
the products of beta functions in (\ref{eq:weightHSC}). These weights can be summed and all possible  
one link loops may be combined into a dual element that plays a similar role as the monomers 
and dimers: They are all local fermionic monomials that can be used to saturate the fermion constraints 
on the sites not occupied by a strong coupling baryon loop.

\begin{figure}
	\includegraphics[width=5cm]{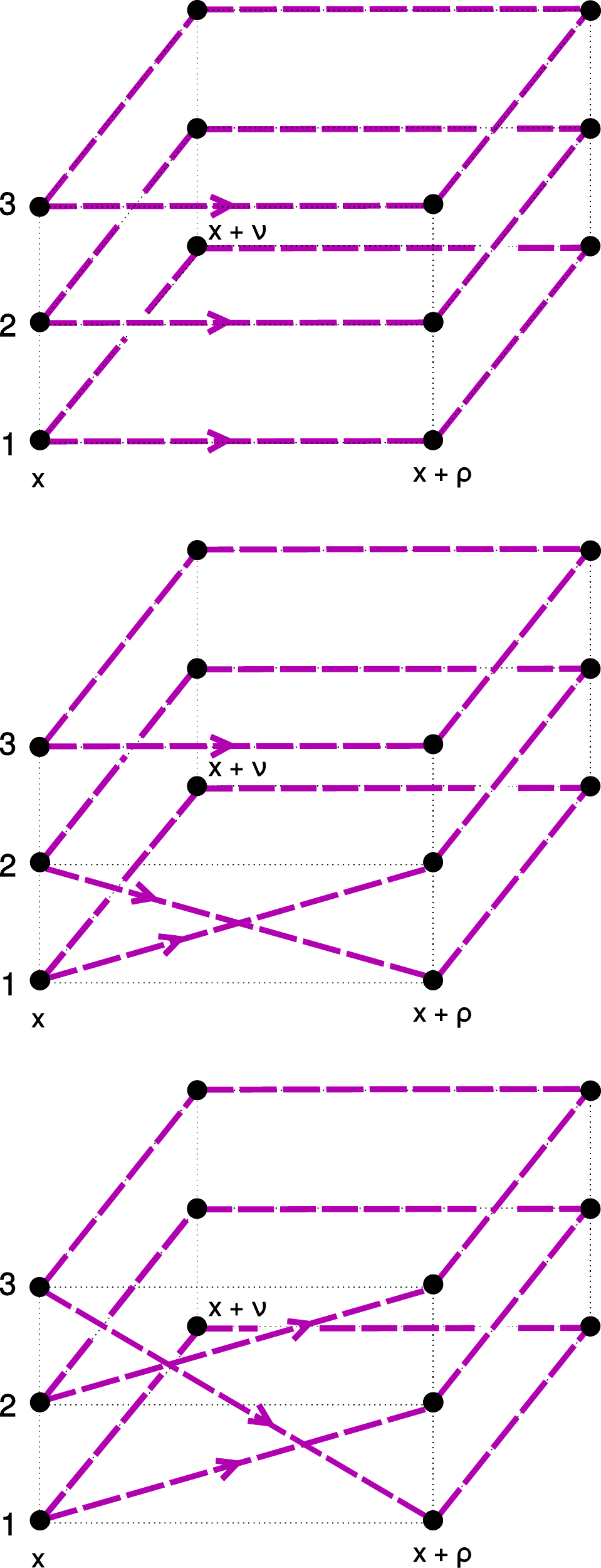}
	\caption{Examples of simple strong coupling baryon loops with different connectivity properties. \label{fig:loop}}
\end{figure}

We now conclude the discussion of strong coupling QCD, by showing that the loop signs (\ref{eq:loop_sign}), 
the gauge signs (\ref{eq:gaugesignSC}) and the constraints (\ref{eq:constraintsSC2A}) --  (\ref{eq:constraintsSC2B})
conspire in such a way, that the remaining strong coupling baryon loops again obey the sign formula 
(\ref{eq:loop_sign}) for staggered fermions. 

For this proof we start with a strong coupling baryon loop made out of only the top left flux elements of 
Fig.~\ref{fig:loop_elements}, i.e., only flux elements with parallel fluxes for all three colors are used. 
A very simple example of such a loop around a single plaquette is shown in the top plot of Fig.~\ref{fig:loop}. 
Obviously such a loop is made out of three copies of the same quark loop. Each one of 
these loops ${\cal L}$ has the sign factor sign(${\cal L}$) for staggered fermion loops as given in 
(\ref{eq:loop_sign}). Thus for the strong coupling baryon loop made out of only the top left flux combinations 
in Fig.~\ref{fig:loop_elements}, we find the sign
\begin{equation}
\sign({\cal L})^3 \; = \; \sign({\cal L}) \; .
\label{baryonsign}
\end{equation}
Note that this identity also holds for temporally winding loops where an additional sign is picked up from the 
antiperiodic boundary conditions.   
Now we can replace the top left flux elements of Fig.~\ref{fig:loop_elements} by one of the other strong coupling baryon  
flux elements where color fluxes cross. An example with an element with one crossing is shown in the middle plot of 
Fig.~\ref{fig:loop} and in the bottom plot we have replaced one of the parallel elements with a strong coupling 
baryon flux with two crossings. However, every crossing of flux also changes the connectivity properties of the loop:
Inserting one crossing either connects two fermion loops into one, see, e.g., the example in the 
middle of  Fig.~\ref{fig:loop}, or splits a loop into two components. Thus every crossing changes the number of 
loops by one, and since every loop comes with an overall minus sign, inserting one crossing changes the fermion 
sign. However, we have shown that the gauge sign in Eq.~(\ref{eq:gaugesignSC2}) changes with the number of
crossings, such that the gauge sign and the fermion sign cancel. Consequently the sign of the baryon loops
is always given by (\ref{baryonsign}), i.e., the signs of the strong coupling baryon loops are the signs for 
loops of a single free staggered fermion. 

We can make the identification of the strong coupling baryon loops with the loops of a free staggered fermion complete
by using the fact  that all the strong coupling flux elements in Fig.~\ref{fig:loop_elements} come with the same weight 1/12.
Thus at every link of the loop we can sum over all six possible fluxes and obtain a total link weight
of 1/2. We conclude that the dual form of strong coupling QCD is a gas of free staggered fermion loops with a 
link weight of 1/2. These loops describe baryons and are embedded in a background of monomers, dimers and 
local link loops, such that the fermion constraints are obeyed.  

We are currently exploring the possibility of updating our form of strong coupling QCD with fermion bags
\cite{Chandrasekharan:2009wc,Chandrasekharan:2012va,Chandrasekharan:2013rpa}: One can 
sum up the weights of all local link loops in Fig.~\ref{fig:link_loops} and all combinations of three dimers that saturate the 
fermion constraints on a single link. All these terms give rise to an effective baryon dimer with a weight 
larger than 1/4, which is the weight of a dimer from staggered fermions with a link factor 1/2. 
Splitting the overall weight of the effective baryon dimer in the form $1/4 + g$, we may
treat the part with factor $g$ as an interaction for the free staggered fermions used for the baryons, together with 
the remaining configurations not yet taken into account, i.e., mixed contributions of monomers and dimers on a link and 
closed chains of single and double dimer links. These interaction terms come with positive factors and can be activated 
according to their weight, such that activated terms delimit the fermion bags inside 
which the free staggered fermions for the 
baryons may propagate.   

\section{Full QCD \label{QCD}}

We complete the presentation of the dual representation in terms of worldlines and worldsheets with discussing the 
case of full QCD. The partition function of full QCD can be written as
\begin{equation}
	\label{eq:partitionQCD}
	Z = \int \! \! D[U] \; Z_{F} [U] \; e^{-S_{G}[U]} \; ,
\end{equation}
i.e., the fermionic partition function $Z_{F} [U]$ given in Eq.~(\ref{eq:partition_fermions}) 
is now integrated over with the Boltzmann factor for the gauge action $S_{G}[U]$ (\ref{eq:actionsu3}).

In Section \ref{QCDatSC} we have obtained the intermediate result 
(\ref{eq:partition}) where the fermionic partition function $Z_{F} [U]$ in a fixed gauge background is already 
expressed as a sum over configurations of the dual fermion variables $s^a_x, k_{x,\nu}^{ab}$ and 
$\overline{k}_{x,\nu}^{ab}$. The dual variables $k_{x,\nu}^{ab}$ and $\overline{k}_{x,\nu}^{ab}$
for fermion hopping activate the corresponding link matrix elements $U_{x,\nu}^{ab}$ and 
$U_{x,\nu}^{ab \, \star}$ which in the strong coupling expression (\ref{eq:partition}) simply were integrated 
over with the gauge field measure $\int \! D[U] = \prod_{x,\nu} \int dU_{x,\nu}$. 

In full QCD the gauge field integral now also has to take into account the gauge field Boltzmann factor, such that
the resulting integral reads 
\begin{equation}
	\label{eq:gauge_integral}
	\bigg[ \prod_{x,\nu} \! \int \!\! dU_{x,\nu} \bigg]  e^{- S_{G}[U]} 
	\prod_{x, \nu} \prod_{a, b} 	\left( U_{x,\nu}^{ab} \right)^{k_{x,\nu}^{ab}}
	\left( U_{x,\nu}^{ab \, \star} \right)^{\overline{k}_{x,\nu}^{ab}} \! .
\end{equation} 
The Boltzmann factor $e^{- S_{G}[U]}$ can again be treated as in Section~\ref{su3gauge}, i.e., 
we expand in abelian color cycles and organize the terms with respect to the links $(x,\nu)$ and 
color indices $a,b$. Thus the remaining gauge field integral reads 
\begin{equation}
	\label{eq:gauge_integral}
	\bigg[ \prod_{x,\nu} \! \int \!\! dU_{x,\nu} \bigg]  
        \prod_{a, b} \left( U_{x,\nu}^{ab} \right)^{N_{x,\nu}^{ab} + k_{x,\nu}^{ab}}
	\left( U_{x,\nu}^{ab \, \star} \right)^{\overline{N}_{x,\nu}^{ab}+\overline{k}_{x,\nu}^{ab}} \! ,
\end{equation} 
where $N_{x,\nu}^{ab}$ and $\overline{N}_{x,\nu}^{ab}$ are the same combinations as defined in 
(\ref{eq:powerNbar2}).
This is the same integral as in the intermediate result (\ref{eq:partitionsumsu33}), only the 
$N_{x,\nu}^{ab}$ and $\overline{N}_{x,\nu}^{ab}$ are now replaced by $N_{x,\nu}^{ab} + k_{x,\nu}^{ab}$ 
and $\overline{N}_{x,\nu}^{ab} + \overline{k}_{x,\nu}^{ab}$. Consequently we can simply follow the steps in 
Section~\ref{su3gauge}. We again write the exponents $U_{x,\nu}^{ab}$ and $U_{x,\nu}^{ab \, \star}$ in the form
\begin{equation}
N_{x,\mu}^{ab} + k_{x,\mu}^{ab} = \frac{Q_{x,\mu}^{ab} + L_{x,\mu}^{ab}}{2} \, , \; 
\overline{N}_{x,\mu}^{ab} + \overline{k}_{x,\mu}^{ab}= \frac{Q_{x,\mu}^{ab} - L_{x,\mu}^{ab}}{2} \, ,
\end{equation}
where 
\begin{equation}
L_{x,\mu}^{ab}  \, = \, J_{x,\mu}^{ab} + K_{x,\mu}^{ab} \; , \; \; 
Q_{x,\mu}^{ab} \, = \, S_{x,\mu}^{ab} + P_{x,\mu}^{ab} \; ,
\end{equation}
where $J_{x,\mu}^{ab}$ and $S_{x,\mu}^{ab}$ defined in (\ref{eq:Jfluxes}) and (\ref{eq:Sfluxes}) collect the 
fluxes and weight arguments for the gauge fields, and $K_{x,\mu}^{ab}$ and $P_{x,\mu}^{ab}$ defined in 
(\ref{eq:kshorthand}) those for the fermions. Again we use $m_{x,\nu}^{ab}$ and $\overline{m}_{x,\nu}^{ab}$
with $(a,b) = (2,1)$, $(2,3)$, $(3,1)$, $(3,3)$ as the auxiliary variables for the binomial decomposition, 
which now run from 0 to $N_{x,\nu}^{ab} + k_{x,\nu}^{ab}$
and $\overline{N}_{x,\nu}^{ab} + \overline{k}_{x,\nu}^{ab}$, respectively. 

Putting things together we find that the dual form of the partition function of full QCD is a sum over configurations
$\sum_{\{s,k,\overline{k},p\}}$ of the fermion dual variables $s^a_x, k_{x,\nu}^{ab}, \overline{k}_{x,\nu}^{ab} \in \{0,1\}$, 
as well as the cycle occupation numbers $p_{x,\mu\nu}^{abcd} \in \mathbb{Z}$,
\begin{equation}
	Z = \!\!\!\!\! \sum_{\{s,k,\overline{k},p\}} \!\!\!\!\!  C_{F}[s,k,\overline{k}] \; W_F[s,k,\overline{k}] \; 
	C_{G}[k,\overline{k},p] \; W_G [k,\overline{k},p]  \, .
	\label{eq:partitionqcd}
\end{equation}
The fermion constraint $C_{F}[s,k,\overline{k}]$ is again given by (\ref{eq:fermion_constraint})
i.e., the admissible worldline configurations are such that every site of the 3-layer lattice is either
occupied by a monomer, is the endpoint of a dimer or is run through by a loop $\mathcal{L}$.
Also the fermion weights $W_F[s,k,\overline{k}]$ are the ones already discussed in (\ref{fermionweight}),
i.e., monomers contribute a factor of $2m$, loops come with a sign $\sign(\mathcal{L})$ given in 
(\ref{eq:loop_sign}) and the chemical potential couples to the temporal winding number $W_\mathcal{L}$ of the loops.

The gauge constraints $C_{G} [k,\overline{k},p]$ are given by 
\begin{eqnarray}
\label{linkconstraintsqcd}
C_{G} [k,\overline{k},p]  & = &  \prod_{x,\mu}  \delta(L_{x,\mu}^{12} + L_{x,\mu}^{13} - L_{x,\mu}^{21} - L_{x,\mu}^{31}) 
\\
&& \hspace{1.2mm} \times  \, 
\delta(L_{x,\mu}^{21} + L_{x,\mu}^{23} - L_{x,\mu}^{12} - L_{x,\mu}^{32}) 
\nonumber \\ 
&& \hspace{1.2mm} \times \, 
\delta(L_{x,\mu}^{11} + L_{x,\mu}^{12} - L_{x,\mu}^{23} - L_{x,\mu}^{33})
\nonumber \\ 
&& \hspace{1.2mm} \times \, 
\delta(L_{x,\mu}^{31} + L_{x,\mu}^{33} - L_{x,\mu}^{12} - L_{x,\mu}^{22})  \; .
\nonumber
\end{eqnarray}
Structurally these are of course the same constraints as for pure gauge theory and strong coupling QCD -- 
after all they are generated by integrating the SU(3) link matrices -- but here in full QCD they link the color flux
contributions from both, the gauge fields via the cycle occupation numbers $p_{x,\mu\nu}^{abcd}$ and the 
fermion loops via $k_{x,\nu}^{ab}$ and $\overline{k}_{x,\nu}^{ab}$. 

The gauge field weights are structurally identical to those of pure gauge theory, but again also the fluxes from the 
fermions contribute through the combined variables $Q_{x,\nu}^{ab}$. The weights are again a sum 
$\sum_{\{l,m,\overline{m}\}}$ over configurations
of the auxiliary plaquette variables $l_{x,\mu\nu}^{abcd} \in \mathbb{N}_0$ and the auxiliary 
binomial variables $m_{x,\nu}^{ab}$ and $\overline{m}_{x,\nu}^{ab}$, 
\begin{widetext}
	\begin{align}
	\label{eq:weight_full}
	& W_G [k,\overline{k},p] \, = \, 2^{\,4V} \!\!\!\!\sum_{\{l,m,\overline{m}\}}  
	\Bigg[ \prod_{x,\mu}  \delta(L_{x,\mu}^{12} - j_{x,\mu}^{21} - j_{x,\mu}^{23} - j_{x,\mu}^{31} - j_{x,\mu}^{33}) \Bigg] \; 
	\Bigg[ \prod_{x,\mu}  (-1)^{L_{x,\nu}^{12} + L_{x,\nu}^{23} + L_{x,\nu}^{31} - j_{x,\nu}^{23} - j_{x,\nu}^{31} }\Bigg]
	\\ \nonumber
         & \hspace{41.5mm} \times \! 
	 \Bigg[ \prod_{x,\mu} \prod_{a = 2,3}  \; 
	 \prod_{b=1,3} \!
	\binom{N_{x,\mu}^{ab} + k_{x,\mu}^{ab}}{m_{x,\mu}^{ab}} \! 
	\binom{\overline{N}_{x,\mu}^{ab} + \overline{k}_{x,\mu}^{ab}}{\overline{m}_{x,\mu}^{ab}} \!
	\Bigg] \,  \Bigg[ \prod_{x,\mu<\nu} \prod_{a,b,c,d} 
	\dfrac{ \left( \beta/2 \right)^{|p_{x,\mu\nu}^{abcd}| + 2\, l_{x,\mu\nu}^{abcd}}}{
	\left(|p_{x,\mu\nu}^{abcd}| + l_{x,\mu\nu}^{abcd}\right)! \; l_{x,\mu\nu}^{abcd}!} \Bigg] 
\\ \nonumber
	& \times \! \Bigg[ \prod_{x,\mu} \B\!\left(\!\dfrac{Q_{x,\mu}^{11} + Q_{x,\mu}^{13} + Q_{x,\mu}^{22} + Q_{x,\mu}^{32}}{2} + 2\right.\!\left|\dfrac{Q_{x,\mu}^{12} + s_{x,\mu}^{21} +s_{x,\mu}^{23} + s_{x,\mu}^{31} + s_{x,\mu}^{33}}{2} + 1\!\right)
	\\
	\nonumber
	& \hspace{7.5mm} \times \B\!\left(\!\dfrac{Q_{x,\mu}^{11} + s^{21}_{x,\nu} + Q_{x,\mu}^{23} - s_{x,\mu}^{23} + s_{x,\mu}^{31} + Q_{x,\mu}^{33} - s_{x,\mu}^{33}}{2} + 1\right.\!\left|\dfrac{Q_{x,\mu}^{13} + Q_{x,\mu}^{21} - s_{x,\mu}^{21} + s_{x,\mu}^{23} + Q_{x,\mu}^{31} - s_{x,\mu}^{31} + s_{x,\mu}^{33}}{2} + 1\!\right)
	\\
	\nonumber
	&\hspace{7.5mm} \times \B\!\left(\!\dfrac{s_{x,\mu}^{21} + Q^{22}_{x,\nu} + s_{x,\mu}^{23} + Q_{x,\mu}^{31} - s_{x,\mu}^{31} + Q_{x,\mu}^{33} - s_{x,\mu}^{33}}{2} + 1\right.\!\left|\dfrac{Q_{x,\mu}^{21} - s_{x,\mu}^{21} + Q_{x,\mu}^{23} - s_{x,\mu}^{23} + s_{x,\mu}^{31} + Q_{x,\mu}^{32} + s_{x,\mu}^{33}}{2} + 1\!\right)\!\! \Bigg],
	\end{align}
\end{widetext}
where we again used the abbreviations $j_{x,\mu}^{ab} \equiv m_{x,\mu}^{ab} - \overline{m}_{x,\mu}^{ab}$ and
$s_{x,\mu}^{ab} \equiv m_{x,\mu}^{ab} + \overline{m}_{x,\mu}^{ab}$ from Eq.~(\ref{eq:binomial}).

We conclude this section on full QCD with addressing two important aspects of the new representation: 
As in the case of pure SU(3) lattice gauge theory, our dual form of the partition sum has the structure of a
strong coupling expansion, and again, our approach allows one to compute all coefficients of this expansion in closed 
form. 

Furthermore, it is obvious how to generalize the construction to several flavors: One simply uses multiple sets
of dual fermion variables, which all couple in the same way to the gauge fields. Thus instead of the variables
$k_{x,\nu}^{ab}$ and $\overline{k}_{x,\nu}^{ab}$ one has flavor sums over such variables and the color fluxes
at each link have contributions from all flavors. These flavor sums over the dual fermion variables  
enter the constraints and weights, which otherwise have the same form as presented in this section. 

\section{Summary and conclusions}

In this paper we have presented a new dual worldline/worldsheet representation of lattice QCD based on the 
abelian color flux approach, where both, the fermion as well as the gauge action are decomposed into minimal 
terms, the abelian color fluxes, that connect different color indices at neighboring sites. After expanding the 
corresponding Boltzmann factors the contributions are organized according to links and the non-abelian
gauge field integrals can be solved in closed form. These integrals lead to weight factors for the fluxes, 
as well as to constraints sitting on the links.

The approach here is presented for three cases: Pure SU(3) lattice gauge theory, strong coupling lattice QCD and
full lattice QCD (the latter two for one flavor of staggered fermions). In the pure SU(3) case the abelian color fluxes are 
generated from abelian color cycles, which are loops in color space closing around plaquettes. The constraints  
restrict the possible configurations of abelian color cycles, and we show that the degrees of freedom responsible for 
long distance physics are closed worldsheets living on a space-time lattice with three layers for the three colors. 

One of the key results is the identification of the constraints of the color fluxes, which are collected in 
Eqs.~(\ref{eq:constraints_final_1}) -- (\ref{eq:constraints_final_4}) and illustrated in Fig.~\ref{fig:constraints_su3}.
The constraints Eqs.~(\ref{eq:constraints_final_1}) -- (\ref{eq:constraints_final_3}) enforce the individual 
conservation of flux for all three colors ({\sl color conservation constraints}), while the constraints 
(\ref{eq:constraints_final_4}) ensure the equal distribution of flux among the colors ({\sl color exchange constraints}). 
All constraints are structurally the same for the three systems studied here, i.e., pure SU(3) gauge theory, 
strong coupling QCD and full QCD, although the sources of color flux are different: Plaquette based
cycle occupation numbers for the worldsheets of gauge degrees of freedom and link based color fluxes for
the fermion worldlines. However, in both cases the constraints (\ref{eq:constraints_final_1}) -- (\ref{eq:constraints_final_4})
are the dual manifestation of the original SU(3) symmetry of the conventional representation. We remark again that the 
dual representation has weight factors with negative signs and a Monte Carlo simulation might be possible only after
finding a suitable resummation scheme.

In strong coupling QCD the abelian color fluxes are generated by loops and dimers of fermions, which together with 
monomers for the mass terms constitute the admissible configurations for the fermions. Again the gauge link integrals
generate constraints for the color fluxes which are structurally identical to the ones in pure SU(3) gauge theory. 
Combining the constraints for the color fluxes with the fermion constraints we show that here the degrees of freedom
that are relevant for long distance physics are strong coupling baryon loops, and the chemical potential 
couples to 3-times their temporal winding number. The interplay of the signs for the  
quark loops with the signs from the SU(3) parameterization conspires to give the baryon loops the same signs as for a 
single staggered fermion. The form we obtain exactly reproduces the strong coupling representation by
Karsch and M\"utter \cite{Karsch:1988zx}. We conjecture that our representation of strong coupling QCD admits a 
fermion bag simulation and we are currently exploring this idea.

Finally, in full QCD the color fluxes on the links receive contributions from the fermion loops and dimers, as well as
the abelian color cycles that represent the gauge degrees of freedom. Consequently also the constraints and the 
gauge weights couple to the combination of these two types of dual degrees of freedom. In full QCD, as well as 
in pure SU(3) gauge theory, the dual representation has the structure of a strong coupling expansion and the abelian
color flux approach allows one to calculate all coefficients of the expansion in closed form. 

We conclude with stressing again that the focus of this work is on analyzing the structure of the constraints in the 
worldline/worldsheet representation, since this provides the manifestation of the original SU(3) gauge symmetry 
in the dual language. The abelian color flux approach, combined with the binomial decomposition of the matrix 
elements that are sums, is a strategy that can be generalized to arbitrary non-abelian gauge groups and several flavors
of fermions. The dual worldline/worldsheet form of such systems of fermions coupled to non-abelian gauge fields 
highlights different properties than the conventional representation in terms of fields, such as the manifestation
of the net-particle number as the temporal winding number of the matter loops. The dual representation 
provides new strategies for further understanding the dynamics of non-abelian 
theories and, e.g., the question how topological properties of the non-abelian gauge fields manifest themselves 
in a worldsheet representation is a problem that will be addressed in the formulation presented here.

\begin{acknowledgments}
	We thank Falk Bruckmann, Philippe de Forcrand, Daniel G\"oschl and Tin Sulejmanpasic for discussions. 
	We acknowledge the support and the hospitality of the MITP in Mainz during the workshop DIMOCA, 
	18-29 September 2017, where part of this work was prepared. Our research is supported by the FWF DK W1203 
	{\sl ''Hadrons in Vacuum, Nuclei and Stars''}, and partly also by the FWF Grant.\ Nr.\ I 2886-N27, as well as the
	DFG TR55, {\sl ''Hadron Properties from Lattice QCD''}.  
\end{acknowledgments}

\bibliography{su3bib.bib}

\end{document}